%% file: main.tex
\documentclass[11pt,aps,prd,reprint,superscriptaddress,citeautoscript]{revtex4-1}
\usepackage[caption=false]{subfig}
\usepackage[version=4]{mhchem}
\usepackage{graphicx,color}
\usepackage[usenames,dvipsnames,table]{xcolor}
\usepackage{placeins}
\usepackage{bm}
\usepackage{amsmath}
\usepackage{amssymb}
\usepackage{mathtools}
\usepackage{siunitx}
\usepackage[super]{nth}
\usepackage{upgreek}
\usepackage{multirow}
\usepackage{array}
\usepackage{color}
\usepackage{gensymb}
\usepackage{longtable}
\usepackage{xfrac}
\usepackage{hyperref}
\hypersetup{colorlinks=true,
            linkcolor=red,
            citecolor=blue,
            urlcolor=orange}
\usepackage{graphicx, type1cm, lettrine}
\newcolumntype{K}[1]{>{\centering\arraybackslash}p{#1}}
\setcitestyle{numbers,super}

\usepackage{ amssymb }
\usepackage{ amsmath }
\usepackage{ amsfonts}
\usepackage{ dsfont }
\usepackage{ amsthm}
\usepackage{algpseudocode, algorithm}
\usepackage{bbm}
\usepackage{ enumerate}
\usepackage{bm}

\newcommand{\beginsupplement}{%
        \setcounter{table}{0}
        \renewcommand{\thetable}{S\arabic{table}}%
        \setcounter{figure}{0}
        \renewcommand{\thefigure}{S\arabic{figure}}%
     }

\makeatletter
\def\maketitle{
\@author@finish
\title@column\titleblock@produce
\suppressfloats[t]}
\makeatother

\AtBeginDocument{\usepackage{booktabs}}

\begin{document}
\title{Autonomous Materials Exploration by Integrating Automated Phase Identification and AI-Assisted Human Reasoning}

\author{Ming-Chiang Chang}
\email{mc2663@cornell.edu}
\affiliation
{Department of Materials Science and Engineering, Cornell University, Ithaca, NY 14853, United States}
\affiliation
{These authors contributed equally to this work}

\author{Maximilian Amsler}
\email{amsler.max@gmail.com}
\affiliation{Department of Materials Science and Engineering, Cornell University, Ithaca, NY 14853, United States}
\affiliation
{These authors contributed equally to this work}

\author{Duncan R. Sutherland}
\affiliation{Department of Materials Science and Engineering, Cornell University, Ithaca, NY 14853, United States}

\author{Sebastian Ament}
\affiliation
{Department of Computer Science, Cornell University, Ithaca, NY 14853, United States}

\author{Katie R. Gann}
\affiliation{Department of Materials Science and Engineering, Cornell University, Ithaca, NY 14853, United States}

\author{Lan Zhou}
\affiliation
{Joint Center for Artificial Photosynthesis, California Institute of Technology, Pasadena, CA 91125}

\author{Louisa M. Smieska}
\affiliation{Cornell High Energy Synchrotron Source, Cornell University, Ithaca, NY 14850, United States}

\author{Arthur R. Woll}
\affiliation{Cornell High Energy Synchrotron Source, Cornell University, Ithaca, NY 14850, United States}

\author{John M. Gregoire}
\affiliation
{Joint Center for Artificial Photosynthesis, California Institute of Technology, Pasadena, CA 91125}

\author{Carla P. Gomes}
\affiliation
{Department of Computer Science, Cornell University, Ithaca, NY 14853, United States}

\author{R. Bruce van Dover}
\affiliation{Department of Materials Science and Engineering, Cornell University, Ithaca, NY 14853, United States}

\author{Michael O. Thompson}
\email{mot1@cornell.edu}
\affiliation{Department of Materials Science and Engineering, Cornell University, Ithaca, NY 14853, United States}

\date{\today}

\begin{abstract}
\input{tex_files/abstract}
\end{abstract}

\maketitle

\section{Introduction}
\input{tex_files/introduction}

\section{Results and Discussion}
\input{tex_files/result}
\section{Conclusion}

\input{tex_files/conclusion}
\section{Methods}
\input{tex_files/methods}

\section{Acknowledgments}\label{sec:ack}
\input{tex_files/acknowledgements}

\section{Author contributions}
\input{tex_files/contributions}

\bibliographystyle{apsrev4-2}
\bibliography{reference}

\clearpage
\onecolumngrid
\section{Supplemental Materials}
\title{Supplementary Information for ``Autonomous Materials Exploration by Integrating Automated Phase Identification and AI-Assisted Human Reasoning''}
\maketitle
\pagebreak
\onecolumngrid
\beginsupplement
\input{tex_files/SI}

\end{document}

%% file: tex_files/abstract.tex
Autonomous experimentation holds the potential to accelerate materials development by combining artificial intelligence (AI) with modular robotic platforms to explore extensive combinatorial chemical and processing spaces.
Such self-driving laboratories can not only increase the throughput of repetitive experiments, but also incorporate human domain expertise to drive the search towards user-defined objectives, including improved materials performance metrics.
We present an autonomous materials synthesis extension to SARA, the Scientific Autonomous Reasoning Agent, utilizing phase information provided by an automated probabilistic phase labeling algorithm to expedite the search for targeted phase regions.
By incorporating human input into an expanded SARA-H (SARA with human-in-the-loop) framework,
we enhance the efficiency of the underlying reasoning process.
Using synthetic benchmarks, we demonstrate the efficiency of our AI implementation and show that the human input can contribute to significant improvement in sampling efficiency.
We conduct experimental active learning campaigns using robotic processing of thin-film samples of several oxide material systems, including \ce{Bi2O3}, \ce{SnO_x}, and \ce{Bi-Ti-O}, using lateral-gradient laser spike annealing to synthesize and kinetically trap metastable phases.
We showcase the utility of human-in-the-loop autonomous experimentation for the \ce{Bi-Ti-O} system, where we identify extensive processing domains that stabilize $\delta$-\ce{Bi_2O_3} and \ce{Bi2Ti2O7}, explore dwell-dependent ternary oxide phase behavior, and provide evidence confirming predictions that cationic substitutional doping of \ce{TiO_2} with Bi inhibits the unfavorable transformation of the metastable anatase to the ground-state rutile phase.
The autonomous methods we have developed enable the discovery of new materials and new understanding of materials synthesis and properties.

%% file: tex_files/introduction.tex
Technological developments to address global challenges are critically dependent on the deployment of new materials with enhanced properties, particularly today in the realm of clean power generation and energy storage~\cite{mirkin_energy_2024,mahlia_review_2014}.
Although the Edisonian method of trial and error has succeeded in the discovery of many materials in the past, the advent of innovative synthesis techniques, accompanied by the added dimension of chemical complexity in the search for new materials, has opened up vast processing spaces that require detailed exploration.
High-throughput (HT) synthesis~\cite{dover_codeposited_2004,hattrick-simpers_perspective_2016} has emerged as a powerful approach to combinatorially map large portions of this search space, supported by accurate \textit{ab-initio} calculations~\cite{kirklin_open_2015, jain_commentary_2013} that predict novel materials using automated computational screening~\cite{kim_machine-learning-accelerated_2018,shen_reflections_2022} along with advanced global optimization algorithms~\cite{oganov_modern_2010, pickard_ab_2011, amsler_crystal_2010}.
The large data sets that can now be produced with such HT methods require equally advanced data analysis techniques to elucidate complex processing-composition-structure-property relationships.
Simultaneously, curated materials information has been accumulated in numerous online repositories accessible for further analysis~\cite{kirklin_open_2015, jain_commentary_2013, zagorac_recent_2019}, marking a significant advancement in data democratization efforts that may also accelerate materials development~\cite{andersen_optimade_2021}.

Data from these direct and \textit{in silico} experiments are also critical as training sets for many machine learning (ML) models, improving predictive models to guide material development.
Prominent recent examples include AlphaFold~\cite{senior2020improved}, which relies on a library of protein data to build models of protein folding, and GNoME~\cite{merchant_scaling_2023}, an inorganic materials generation framework based on graph neural networks trained on materials data.
While pre-trained ML models like deep neural networks and foundation models~\cite{choudhary2022recent, takeda_foundation_2023, batatia2023foundation} excel at interpolation and can work well in the presence of sufficiently large data, active learning (AL) approaches are better suited to guide an exploratory process to gain insight when existing data are sparse and hence are more useful for accelerating material experiments~\cite{settles_active_2009}.
AL algorithms aim to successively improve models (ML or others) by proposing optimal experiments that provide the most useful information given existing knowledge, thereby driving the learning process effectively with a feedback loop and ensuring data- and time-efficiency.
Successful demonstrations of AL in materials discovery have been reported for several applications, including phase-change materials~\cite{kusne_on-the-fly_2020}, NiTi-based shape memory alloys with low thermal hysteresis~\cite{xue_accelerated_2016}, and selective synthesis of carbon nanotubes~\cite{nikolaev_discovery_2014,nikolaev_autonomy_2016}.
Meanwhile, several autonomous robotic laboratories for closed-loop synthesis of novel organic compounds have been established~\cite{li_synthesis_2015,kitson_digitization_2018,christensen_data-science_2021,manzano_autonomous_2022,strieth-kalthoff_delocalized_2023}, while analogous infrastructures for inorganic materials synthesis are still under development~\cite{szymanski_autonomous_2023,wang_challenges_2024,leeman_challenges_2024}. 

Despite the enormous potential that AI offers for autonomous and efficient experimental planning in high-dimensional space~\cite{tabor_accelerating_2018,suh_evolving_2020,pyzer-knapp_accelerating_2022}, physical reasoning with experimental results remains a challenge for AI.
In contrast, humans struggle to understand high-dimensional distributions but are able to interpret results using intricate physical intuition and past experiences to guide experiments.
For example, when humans observe unexpected behavior, they often significantly alter what had been identified as the next experiment.
Ideally, experimental workflows should leverage the unique advantages of AI and humans:
Autonomous AI agents can readily relieve scientists from the details of experiment planning, allowing them to instead focus on drawing physical meaning from the data in real-time.
This is akin to AI assistants, such as agentic coding in the programming domain, where the AI assists in writing code for very specific functions while engineers concentrate on the overall code integration and high-level tasks~\cite{moradi_dakhel_github_2023,peng_impact_2023,yetistiren_evaluating_2023}.
The critical question is where the appropriate boundary should be established, and how to create an effective interface to facilitate a complex, well-orchestrated interplay between humans and machines, ultimately optimizing the ``human-in-the-loop” AI for complex material discoveries~\cite{mosqueira-rey_human---loop_2023,adams_human---loop_2023}.

To this end, we expand the capabilities of SARA, the Scientific Autonomous Reasoning Agent~\cite{sara2020vision}, with an enhanced workflow that incorporates human scientist reasoning that directs evolving objectives within an autonomous experimental campaign.
In particular, we demonstrate ``SARA with human-in-the-loop'' (SARA-H) in campaigns that integrate autonomous high-throughput processing with lateral-gradient laser spike annealing (lg-LSA)~\cite{bell_lateral_2016} and real-time structural characterization~\cite{chang2023probabilistic} of thin films.
As the knowledge of the processing-composition landscape develops, the human in the loop alters priorities, moving from exploration to optimizing synthesis conditions for targeted phases (exploitation).
The advantage of incorporating phase information and advanced acquisition functions is demonstrated in both synthetic tests and AL experiments conducted on three complex oxide systems, \ce{Bi2O3}, \ce{SnO_x} suboxides, and the pseudo-binary \ce{Bi-Ti-O} system. 
We discuss our findings for \ce{Bi_2O_3} and \ce{Bi-Ti-O} extensively in Sects.~\ref{sec:bi2o3} and~\ref{sec:bitio}.
The \ce{SnO_x} suboxide system serves as a reference benchmark and the details are  included in the Supplemental Materials~\cite{SI}.
By incorporating human expert interaction within the AL cycles, we enable the targeted synthesis of specific material phases.
For example, within the \ce{Bi-Ti-O} system, 
we targeted and mapped the phase field of bismuth-substituted anatase,
a phase that has been predicted to exhibit enhanced photocatalytic behavior~\cite{satoh_metastability_2013,cui_first-principles_2016,vu_anataserutile_2012}.
With these advancements, SARA-H represents a significant milestone in the development of accelerated experimental materials discovery methods.

%% file: tex_files/result.tex
\subsection{Targeted Material Synthesis with SARA-H\label{sec:workflow}}

\begin{figure*}[ht!]
  \includegraphics[width=\linewidth]{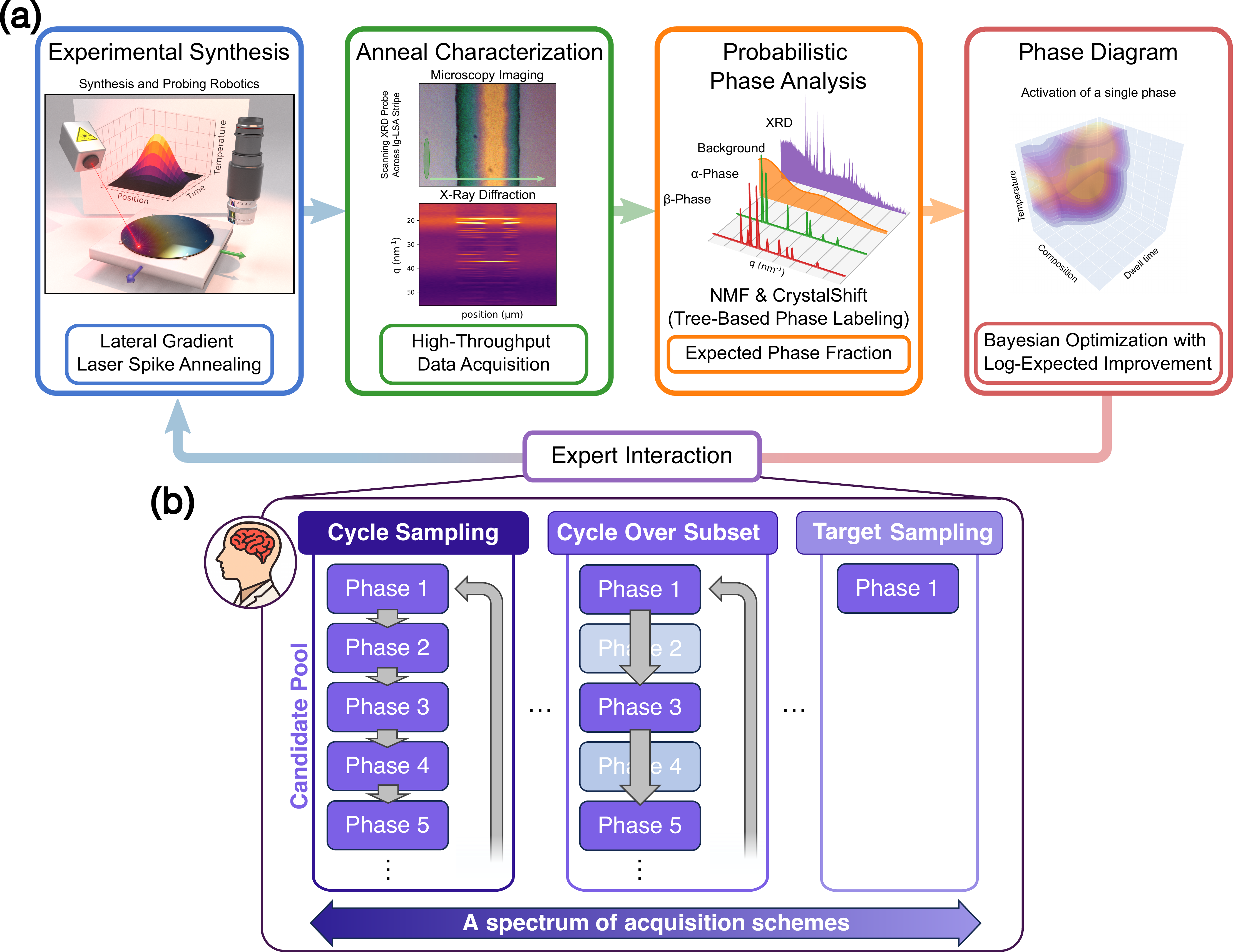}
  \caption{The overall workflow of SARA-H.
  (a) The autonomous experimental workflow or SARA-H.
  The leftmost panel represents SARA-H's synthesis technique, which is based on the annealing of amorphous thin film samples with lg-LSA under specific conditions.
  A 3D-rendered experimental setup is shown, with the laser to the left, a camera to the right representing characterization and the power profile in the backdrop, and the thin-film sample mounted on a stage.
  An annealed lg-LSA stripe (second panel) is then characterized with narrow-beam synchrotron X-ray diffraction.
  The top image shows a micrograph of an annealed stripe, together with the nominal XRD footprint that is scanned across the stripe, while the bottom image presents the corresponding XRD heat map.
  The third panel illustrates analysis of the diffraction data on-the-fly using the probabilistic muiltiphase-labeling algorithm CrystalShift.
  Finally, the expected phase fraction from the probabilistic phase model can be utilized to construct a phase diagram (shown as a probability distribution in a high dimensional processing space), which is progressively improved using Bayesian optimization.
  The AL loop is closed by performing the next, optimal synthesis.
  (b) The optimization strategy may be augmented with human expertise to further enhance the targeted synthesis of specific phases.
  For a given material system, the expert prepares a list of candidate phases to study in the active learning run.
  During the run, the expert monitors results and determines which candidate phases to retain in the the active learning loop for optimization.
  This human-in-the-loop behavior creates an interface for experts to control the behavior of the autonomous experimentation.
  Details on the time scales of every step involved in the SARA-H workflow are compiled in Sec.~\ref{sec:timescales} of the Supplemental Materials.}
  \label{fig:workflow}
\end{figure*}

The SARA-H framework was designed to efficiently explore synthesis phase diagrams and to target the synthesis of specific phases.
We particularly target metal oxides that can form numerous metastable polymorphs under ultrafast annealing conditions achieved with lg-LSA.
Typically, the conditions under which phases can be synthesized are constrained by the precursors, stoichiometry, synthesis methods, and/or limited ranges of thermodynamic processing quantities including temperature and pressure.
Even if synthesized and quenched to ambient conditions, many phases often decompose into more stable (but less desirable) states upon reheating or under operating temperatures in relevant device applications.
Therefore, significant efforts have been made to better understand the thermodynamic scale of metastability in inorganic materials~\cite{sun_thermodynamic_2016, amsler_exploring_2018} and to develop strategies to stabilize them under operating conditions after formation~\cite{hanaor_review_2011}.

The general workflow of SARA-H is shown in Fig.~\ref{fig:workflow}.
Our experimental synthesis module processes thin film compositional libraries, which were previously prepared by depositing precursor amorphous thin films using radio-frequency (RF) magnetron sputtering onto heavily doped silicon wafers with a thermally grown silicon oxide buffer layer.
Composition spreads can be deposited by co-sputtering multiple targets to achieve a smoothly varying compositional gradient across the wafer, with the stoichiometry being a function of position on the wafer, $\textbf{c}(x, y)$.

The processing of these thin-film libraries was conducted using lg-LSA, which forms both equilibrium and metastable phases at high temperatures, and kinetically traps them to ambient conditions by rapid heating and quenching.
In contrast to other traditional annealing methods for thin film samples, including hot plates, furnaces, or rapid thermal annealing~\cite{Borisenko1997}, lg-LSA enables controlled rapid processing in spatially confined regions, with quench rates of $10^4-10^7$~K/s and peak temperatures $T_p$ limited only by the melting point of the substrate (1410~$^\circ$C for Si substrates).
The scanned laser has an intentional lateral intensity profile (typically near Gaussian), giving each lg-LSA stripe a spatially varying thermal profile.
Process temperature and time are dependent on the laser power and the scan velocity, which are characterized by a peak temperature at the center of the stripe $T_c$, and a dwell time related to the heating and quench rates.
For the experiments in this work, the full width at half-maximum (FWHM) of the temperature distribution was $x_w\approx 360$~$\mu$m across the stripe, with some variation based on the anneal condition.
The inset in the \textit{Experimental Synthesis} panel of Fig.~\ref{fig:workflow}(a) is a schematic illustrating the annealing temperature distribution during lg-LSA.
The heating durations and the associated heating/quench rates are characterized by a dwell time $\tau$ defined as the FWHM of the laser spot divided by the stage velocity.
The cooling/quench rate, in particular, can be extremely high (up to $10^7$~K/s) due to direct contact of the thin film with a thermally conductive Si substrate.
Each lg-LSA stripe has a common dwell $\tau$, but a continuous range of temperature conditions across the stripe, yielding multiple phases as a function of the peak annealing temperature.

Following each experimental lg-LSA stripe anneal, the resulting phase behavior was analyzed using multiple appropriate characterization techniques, as illustrated by the \textit{Anneal Characterization} panel in Fig.~\ref{fig:workflow}(a).
In previous work~\cite{ament_autonomous_2021}, we employed microscopy imaging and optical spectrometry to identify structural phase changes that were directly correlated with alterations in the optical properties of oxide films~\cite{sutherland_optical_2020}.
These optical measurements across each lg-LSA stripe, were subsequently used to delineate phase boundaries without explicit crystallographic phase identification, resulting in the creation of an unlabeled processing phase diagram. 
In the current work, we modified analysis agents in SARA-H to incorporate powder X-ray diffraction (XRD) to directly inform the agents of specific crystalline phases synthesized during the process, as shown in the \textit{Probabilistic Phase Analysis} panel in Fig.~\ref{fig:workflow}(a).
To achieve this, a focused 20~$\mu$m wide X-ray beam was scanned across each stripe, generating diffraction patterns as a function of lateral location (associated with the temperature gradient).
The complete diffraction data of an annealed stripe were fed into a nonnegative matrix factorization (NMF) algorithm~\cite{kumar2013fast} to reduce the dimensionality and accelerate data processing.
This specific NMF algorithm identified the most representative XRD patterns that could be used to reconstruct the full XRD data matrix with minimal error, thereby ensuring that it did not hallucinate artificial XRD patterns.
The identified XRD bases were fed into CrystalShift~\cite{chang2023probabilistic}, a tree-based probabilistic multiphase labeling algorithm that rapidly analyzes the XRD patterns.
The labeling results were then combined with the linear coefficients of the NMF to provide an expected phase activation distribution along with their related uncertainties as a function of temperature (for the given dwell).
This expected phase activation serves as a robust proxy for quantifying the phase fraction as illustrated by the \textit{Phase Diagram} panel in Fig.~\ref{fig:workflow}(a).

The expected phase activation of the annealed stripes was used to create and continuously update a Gaussian process (GP) model of a phase map within the parameters \{$\textbf{c}, \tau, T_p$\}, which encodes the fraction $p_k$ of each present phase $k$.
This model not only aids in constructing an accurate phase diagram, but also doubles as the surrogate model for optimizing the acquisition functions for AL cycles, which determines the next anneal based on the available information.
Although the previous version of SARA was limited to optimizing the phase-boundary map by targeting maximal gradient regions in optical data~\cite{ament_autonomous_2021}, the probabilistic phase distribution produced by XRD and CrystalShift allows more refined objectives.
In particular, we can employ Bayesian optimization to perform targeted materials synthesis.
At the core of such optimizations, we utilized a recently developed acquisition function based on expected improvement (EI) called logEI~\cite{ament2023unexpected}, which prevents numerical underflow and is therefore more stable and easier to optimize than the regular EI function; this function then served as the foundation for our efficient sampling strategies. 
During both surrogate construction and the optimization of the acquisition function, proper propagation of experimental errors was included to increase the robustness and performance of the AL agent.
A comprehensive benchmark of different sampling methods is discussed in Sec.~\ref{sec:synthetictests}.

Finally, we incorporate human-machine interaction with SARA-H using a simple yet powerful interface, represented by the \textit{Expert Interaction} panel in Fig.~\ref{fig:workflow}(a).
The advantage of human scientists involved in the autonomous experiment is that they can provide the model with higher-level, non-deterministic information that enhances the decision-making process based on reviewing the current state of the model and the aggregated data.
The types of expert knowledge can be diverse, ranging from visually identifying the melting point or delamination conditions of a specific phase, to constraining the search boundaries of the model or correcting misclassifications of XRD data.

In this work, human experts can modify the objective of the AL cycle on-the-fly.
At the start of an AL trial, the analysis agent is typically provided with a complete list of potential phases in the target material system, shown as the \textit{Cycle Sampling} tab in Fig. \ref{fig:workflow}(b).
The AL cycles through this list and attempts to maximize the expected phase fraction of one targeted phase in each loop.
The human can examine the data during the run and decide to add or remove candidate phase(s) (illustrated by the \textit{Cycle Over Subset} tab in Fig. \ref{fig:workflow}(b)).
The reasons for removing candidate phases may include, but are not limited to, the sufficient convergence of the phase field for a specific phase, the likelihood that the phase cannot be stabilized, or a greater interest in certain phases over others.
This round-robin fashion cycle sampling method prevents the expensive evaluation of a multivariate Gaussian process required for optimizing entropy-like objectives.

This implementation of 
SARA-H is also capable of mimicking a scientist's behavior when searching for a specific phase, and attempting to optimize the conversion rate of the same phase for every loop (denoted as \textit{Target Sampling} in Fig.~\ref{fig:workflow}(b)).
Our interface enables a spectrum of acquisition schemes across the exploration-exploitation extrema by adjusting candidate phases, allowing real-time control over autonomous experimental behavior.
In particular, we incorporate this human-AI collaboration scheme in the autonomous experimental process for the complex pseudo-binary Bi-Ti-O system (discussed in detail in Sec.~\ref{sec:bitio}).

\subsection{Synthetic Benchmarks for Sampling Strategies\label{sec:synthetictests}}
The task of identifying the optimal synthesis conditions for a specific phase differs from the common problems typically used to benchmark AL methods, such as the Ackley function~\cite{ackley2012connectionist}, which are often designed to have challenging aspects for AL (e.g., multiple local minima, narrow valleys, flat loss contours, etc.).
In efforts to identify the experimental condition that contributes to phase formation rates, the target distribution is generally assumed to be straightforward, featuring a well-defined maximum and non-zero values contained within a single continuous volume in the parameter space, with few exceptions (phases) that may include more than one continuous volume.
To quantify the benefits of using phase information to solve this specific problem and to statistically compare different sampling strategies, we conducted tests on an experimental dataset in a three-dimensional parameter space, where the first two degrees of freedom (DOF) represent the peak anneal temperature and dwell time, and the third DOF represents a compositional variable. 
The system contains 8 phase regions that were fitted from real experimental data obtained for Bi-Ti-O, a system that will be discussed in more detail in Sec.~\ref{sec:bitio}.

We assessed the efficiency of various sampling strategies primarily using two benchmark metrics: the median and the interquartile range (IQR) of regret $r$, where $r$ is a measure that expresses the relative difference between the maximum phase activation of phase $k$ found by the AL model, $a_k(\bm{\theta}^*)$, and the ground truth maximum phase activation, $a_k(\hat{\bm{\theta}})$, under a given set of parameters $\bm{\theta}=(\textbf{c}, \tau, T)$. 
The median regret is its median for all test trials at a specific iteration. 
We benchmarked four different sampling strategies, including random sampling and uncertainty sampling as baseline methods, and the two methods proposed in this work, namely, target sampling and cycle sampling (results for pseudo-random methods are shown in Fig.~\ref{fig:pseudorandom} of the Supplemental Materials~\cite{SI}).

For each of the above AL sampling strategies, their stripe acquisition variants were also tested, a method developed and introduced in our previous work~\cite{ament_autonomous_2021}.
The stripe acquisition considers a temperature \textit{range} of an lg-LSA stripe to identify the optimal conditions for the subsequent anneal, which can be considered a form of batch processing.
Benchmarking the stripe variants evaluates the extent to which the temperature-gradient phase information in lg-LSA stripes enhances the performance of our workflow.
500 synthetic AL trials were conducted for each acquisition scheme to acquire performance statistics.
The stripe variants were repeated 100 times due to their longer run times.

\begin{figure*}
  \includegraphics[width=0.85\linewidth]{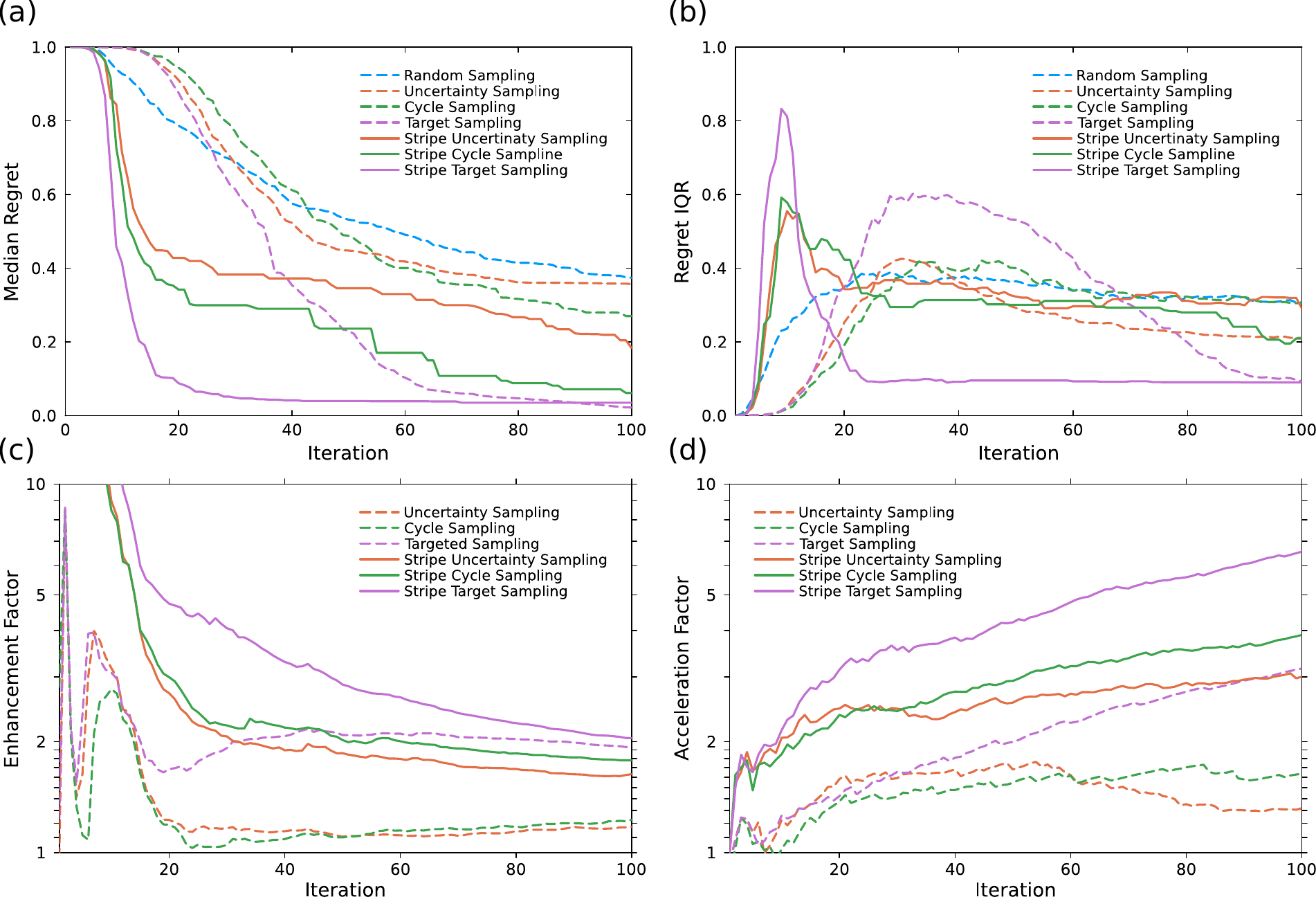}
  \caption{Performance of various active learning sampling strategies in a synthetic phase space.
  (a) The median regret on one of the phases for a materials system containing a total of 8 phases against the number of iterations with different sampling strategies.
  Solid lines are the stripe variant while the dashed lines represent the single-point counterpart.
  (b) The IQR of the regret versus the number of iterations for each sampling strategy, for the same phase as in (a).
  Compared to random sampling, the regret IQR for other methods rises and decreases earlier and more sharply. This indicates the efficient exploration and exploitation of active learning methods.
  (c) The enhancement factors represent the ratio between the objective values obtained from AL and those achieved through random sampling at various stages, averaged over all 8 phases present in the materials system. This quantity illustrates the degree to which the results improve when utilizing AL techniques. 
  (d) The acceleration factors represent the ratios of the average iterations needed to achieve the same objective value for random sampling with respect to each AL strategy,  averaged over all 8 phases that exist.}
  \label{fig:benchmark}
\end{figure*}

Fig.~\ref{fig:benchmark}(a) shows the representative median regret result for one of the eight targeted phases.
Our results clearly demonstrate that the use of AL techniques significantly reduces regret compared to random sampling, although not all of these techniques perform equally well.

The uncertainty sampling strategy served as a baseline for using the active learning scheme.
It selects the condition for the next AL iteration that maximally reduces the overall uncertainty of the model without knowing any phase information.
Still, it outperforms random sampling in the long run but its performance plateaus due to lack of an ability to exploit deeper understanding of the system.

Target sampling employs the logEI acquisition function to maximize the expected phase fraction of a single, targeted phase.
This allows the AL to hone in on one specific phase and is, by design, better suited to exploit phase information compared to uncertainty sampling, which samples the entire parameter space more uniformly.

In contrast, cycle sampling iterates through all possible candidate phases (twelve in total, with eight ground-truth phases and four distractor phases), with the aim of optimizing the activation of a single phase at each AL iteration.
This strategy mimics a scientist who has a list of potentially interesting phases but is not yet certain which phase or subset of phases to explore in greater detail.
Although we found this strategy to be less efficient for identifying an optimal synthesis condition for one specific phase compared to uncertainty sampling in the early stages of an AL campaign, it generally outperformed uncertainty sampling after several AL iterations, showing the gain from incorporating phase information into decision-making.
The utilization and exploitation of phase-aggregated information is particularly evident in cycle sampling, as clear steps reducing the regret can be observed in the dashed green line.
The iterations at which these steps occur correspond to the iterations at which the phase plotted in Fig.~\ref{fig:benchmark}(a) is the target phase.

All sampling strategies benefit significantly when the stripe acquisition method was incorporated, illustrated by the difference between the dashed and solid lines of the same color in Fig.~\ref{fig:benchmark}(a).
This demonstrates effective integration of the high-throughput characterization of lg-LSA into the acquisition function.
It is an example of the general value of incorporating dense and relevant information from high-throughput experiments into AL.

Next, we investigated the IQR of regret to better understand the trade-off between exploration and exploitation in the various AL strategies, as shown in Fig.~\ref{fig:benchmark}(b) for the same targeted phase as in Fig.~\ref{fig:benchmark}(a).
The IQR is defined as the difference between the \nth{75} and \nth{25} percentiles of the regret distribution in the statistical ensemble, and can be used as a proxy of the behavior of AL strategies.
Large IQR values arise when each statistical AL run samples different regions of the parameter space (exploration), while small IQR values indicate similar behaviors among each AL instance (exploitation). 
In Fig.~\ref{fig:benchmark}(b), all IQR curves for every sampling strategy follow a similar trend: initially, the IQR increases, then experiences a downturn, and eventually flattens out gradually.
These three domains can be interpreted as follows:

First, the initial increase in the IQR indicates that the AL is exploring the space, which leads to significantly different maximum phase activation observed for each AL trial.
The steeper this increase, the more efficient the exploration phase.
Stripe sampling variants are particularly efficient since more information is collected from each stripe.

Second, after the exploration stage, all methods experience a downturn in IQR within the second domain.
The sources for this decrease are two-fold: (i) it may be caused by the exploitation of information by the AL runs.
The drop in IQR indicates that AL instances have explored enough of the parameter space to take advantage of the information to optimize the phase and have begun to reach a consensus on where the target phase should be with a final, ultimately similar $a_k(\bm{\theta}^*)$, which leads to a smaller IQR.
(ii) There is an effect of overlapping information in the newly sampled point with the previous one, which is, in fact, the only source that leads to a lower IQR in random sampling.
The steeper the drop in IQR, the more efficiently the AL is exploiting the phase-space information.
Using the above interpretation, we observe that the IQR curves of cycling and target sampling overlap with that of uncertainty sampling at early iterations, as they are mathematically equivalent before any significant phase fraction has been discovered.
However, after the IQR starts to decrease, the target sampling shows a much larger negative slope.
This observation, along with lower regret compared to uncertainty sampling, demonstrates that target sampling strategies exhibit markedly faster learning compared to uncertainty or random sampling.

Third, within the last domain, the IQR curves plateau at a residual value.
A large residual IQR indicates that additional AL instances do not converge to the same final model.
A small residual spread is critical as it provides evidence for both efficient learning and robust model convergence. Note that a similar behavior is observed for the IQR curves of all 8 phases in our benchmark set, as shown in Fig.~\ref{fig:fulliqr} of the Supplemental Materials~\cite{SI}.

Overall, our synthetic benchmarks demonstrate that the target and cycle sampling strategies perform best, significantly outpacing both random sampling and uncertainty sampling.
Note that if the candidate set is reduced due to human intervention, the sampling performance is expected to be between the two extrema, namely the target sampling and the cycle sampling.

To quantify the benefit of using AL, the enhancement and acceleration in the maximum activation found were benchmarked and are shown in Fig.~\ref{fig:benchmark}(c) and  Fig.~\ref{fig:benchmark}(d), respectively, using the metric proposed in Liang \textit{et al}~\cite{liang2021benchmarking}.
The results show that AL techniques can accelerate the process of finding the best condition for synthesis of a specific phase by a factor of two to five, and can find a conversion rate that is roughly two times higher than random sampling. Further discussions on both the enhancement and acceleration factors are provided in Sec.~\ref{sec:enhancement-acceleration} of the Supplemental Materials~\cite{SI}.

\subsection{Active Learning Experiments in Real Materials Systems}

\subsubsection{Active Learning on Unary Systems}
\label{sec:bi2o3}
Having established the effectiveness of cycle sampling based on synthetic tests, we turn our attention to the synthesis of real materials to explore phase formation in complex oxides. 
We first considered the \ce{Bi2O3} system, which has a phase diagram encompassing at least five distinct experimentally observed polymorphs~\cite{harwig_structure_1978,cornei_new_2006}.
The metastable $\delta$-\ce{Bi2O3} phase has attracted considerable attention, especially in the context of its application as a solid electrolyte in fuel cells~\cite{shuk_oxide_1996}.
The defective fluorite-type crystal structure of the $\delta$-\ce{Bi2O3} phase is characterized by a high concentration of oxygen vacancies, leading to the highest oxygen ion conductivity among solid oxides documented to date~\cite{fruth_structural_2007}.
Its application is hindered by a narrow thermodynamic stability window, limited to temperatures between $727 - 824~^\circ$C, preventing its use on an industrial scale.
Attempts to stabilize $\delta$-\ce{Bi2O3} at room temperature through yttrium~\cite{takahashi_high_1975} or rare earth metal substitutions  result in compromised ion conductivity.
Consequently, ongoing efforts focus on identifying pathways to maintain phase-pure $\delta$-\ce{Bi2O3} at ambient conditions using grain boundary engineering~\cite{jeong_physically_2022} and sophisticated thermal processing~\cite{bell_rapid_2021}.

We previously utilized SARA to autonomously explore the phase map of $\delta$-\ce{Bi2O3} using lg-LSA~\cite{ament_autonomous_2021}, identifying the boundaries of the sought-after $\delta$-\ce{Bi2O3} within the processing phase diagram using purely optical techniques, including microscopy and reflectance spectrometry.
In this study, we re-investigated this system, now constructing a fully labeled phase diagram on-the-fly using CrystalShift's capability to rapidly disentangle XRD patterns.
The \ce{Bi2O3} system is particularly well-suited as a use case for our new algorithm, allowing for the comparison of the resulting phase diagrams.

For SARA-H's closed-loop experiment, we began with a pristine thin film of amorphous \ce{Bi2O3}, deposited under the same conditions as in our prior work~\cite{ament_autonomous_2021}.
The AL agents were initialized with simulated XRD patterns derived from a selection of atomic structures of the potential candidate phases, specifically the $\alpha$-, $\beta$-, $\gamma$-, $\delta$-, and $\epsilon$-phases of \ce{Bi2O3} retrieved from ICSD~\cite{zagorac_recent_2019}, utilizing cyclic sampling for all phases in the candidate pool.
As described in detail in Sec.~\ref{sec:synthetictests}, this sampling technique aims to iteratively maximize the phase activation of a phase $k\in \{\alpha, \beta, \gamma, \delta, \epsilon\}$, using the logEI algorithm to determine the next experiment~\cite{ament2023unexpected}.
In SARA-H, the relevant phase $k$ is cycled over the candidate phases at each AL iteration; hence, in every step, a different phase is targeted.
Note that although only one phase is targeted at each iteration, the knowledge of \text{all} candidate phases is updated (i.e., the present phase fraction at a given processing condition).
Therefore, the uncertainty in the phase map is iteratively reduced over \text{all} candidate phases simultaneously.

\begin{figure*}
    \includegraphics[width=0.85\linewidth]{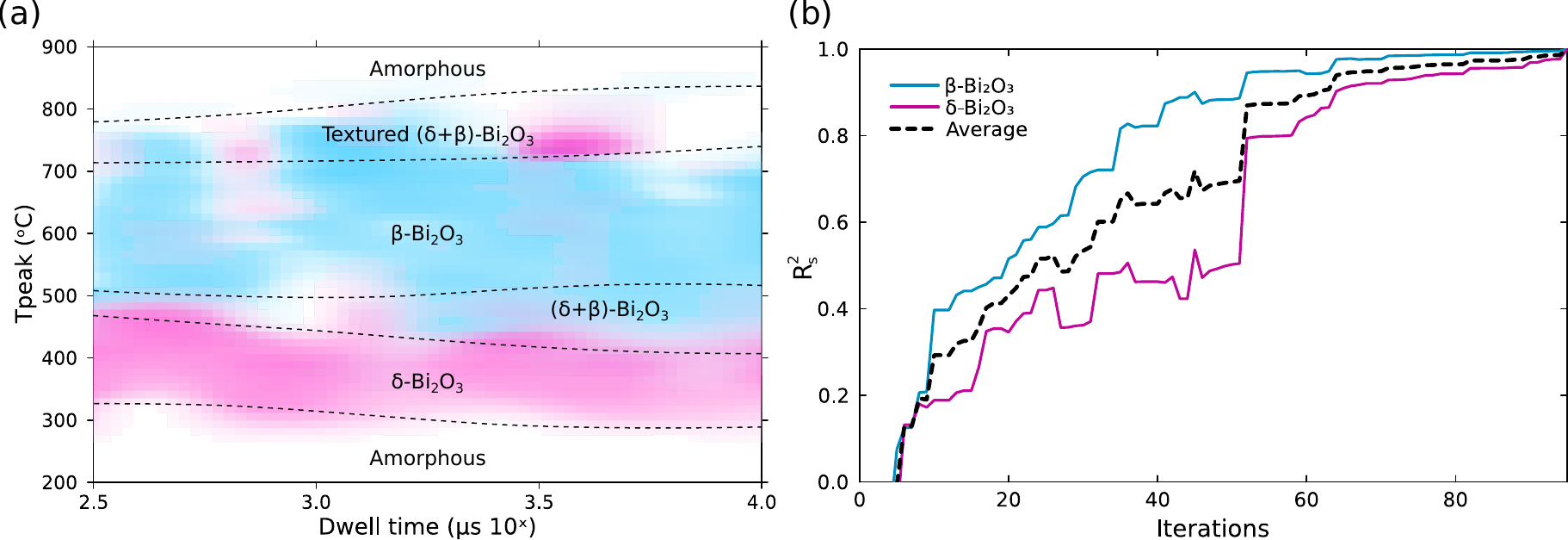}
  \caption{ (a)  Processing phase diagrams of the \ce{Bi2O3} system after SARA-H's converged AL cycles with 95 iterations. Blue and purple domains represent the presence of $\beta$-\ce{Bi2O3} and  $\delta$-\ce{Bi2O3}, respectively. The saturation of each region is proportional to the corresponding phase activation with amorphous domain represented as uncolored regions. The dashed black lines approximate the phase boundaries and are intended  as a guide for the eye.  (b)  $R_s^2$ convergence as a function of the AL iterations for each of the two observed phases, with their average shown as the dashed line.}
  \label{fig:bi2o3}
\end{figure*}

Fig.~\ref{fig:bi2o3}(a) presents the final, labeled phase diagram of \ce{Bi2O3} after 50 iterations.
This phase diagram is derived from expected phase fractions, using color saturation to indicate the relative abundance of each phase.
A comparison with panel D in Fig.~3 of Ref.~\onlinecite{ament_autonomous_2021} shows good agreement between the phase maps.
There is a temperature shift compared to previous results due to the improvement in the temperature calibration method (see Sec.~\ref{sec:lglsa}).
The Bi-O system is particularly challenging for automatic phase labeling algorithms as the XRD patterns of $\beta$-\ce{Bi2O3} are very similar to $\delta$-\ce{Bi2O3} and differ only by some peak splitting due to symmetry breaking.
This similarity makes determining the coexistence region through XRD quite difficult.
However, the probabilistic labeling helps to identify the two-phase region, leading to creation of a reliable phase diagram.

Note that the crystallographic textured ($\delta+\beta$)-\ce{Bi2O3} observed above approximately 700~$^\circ$C arises from the melt-quenching process.
The pronounced texturing results in the absence of many peaks from the XRD data, rendering the NMF and phase labeling less reliable.
Consequently, assistance from a human expert is required to clearly identify such regions.
This can be accomplished either on-the-fly during the AL campaign or, as in our case, as a postprocessing step while analyzing the data to produce the final phase diagram.
While dealing with XRD data from such textured films may occasionally produce erroneous labeling within our workflow, we rarely expect significant texturing or the formation of grain sizes comparable to the X-ray probe size, both of which can lead to missing peaks in XRD patterns before the melting temperature is reached.
On the other hand, if texturing conditions are expected \textit{a priori}, it is possible to reliably characterize these features on-the-fly by incorporating the mediation technique outlined in the original work on CrystalShift~\cite{chang2023probabilistic}.

The learning behavior of SARA-H's AL agent is shown in Fig.~\ref{fig:bi2o3}(b), where the $R^2_s$ (see Sec \ref{sec:metrics}) score at each iteration is plotted in relation to the fully converged phase diagram constructed from the collected XRD data from 95 anneal experiments.
The figure shows that after only some 50 iterations, the average $R^2_s$ score reaches a value exceeding $0.8$, indicating that the phase diagram has already converged.
A similar value of $R^2_s$ was achieved after more than 70 iterations in Ref.~\onlinecite{ament_autonomous_2021} for the same materials systems, which is 40\% more than what is observed in this work.

The reason for this faster learning can be attributed to two key factors: (i) Accelerated learning is facilitated by the increased richness of physical information contained in the XRD data compared to reflectance spectrometry.
Rather than relying solely on the rate of change in the optical data, the XRD provides phases information at different conditions, allowing direct optimization of target phases.
(ii) The objective of maximizing phase activations is significantly easier to optimize than focusing solely on phase boundaries based on optical gradients, as the phase space occupied by boundary regions constitutes only a fraction of the phase fields themselves.
This issue is further exacerbated when moving towards systems with increasing compositional degrees of freedom, as the likelihood of sampling a phase boundary region compared to a phase field diminishes further.

\begin{figure*}[ht!]
     \centering
         \includegraphics[width=0.9\linewidth]{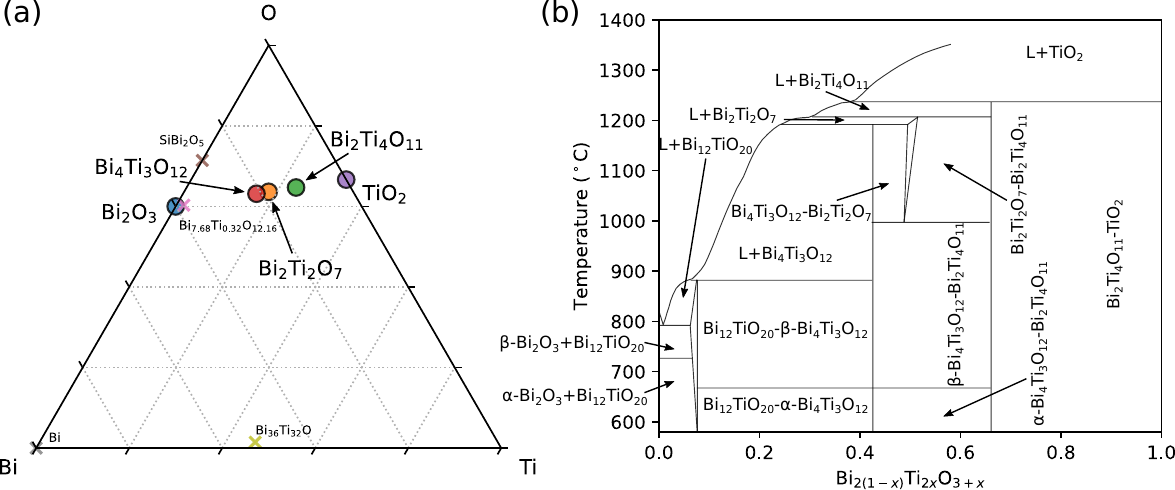}
         \caption{ (a) Gibbs triangle of the ternary phase space Bi-Ti-O. The diagram features all compositions considered in the candidate pool for our probabilistic phase determination algorithm, represented as either circles or crosses. Circles indicate compositions that were observed in the experiment while crosses denote phases that did not form. Note that most candidate compositions lie along the tie line between \ce{Bi2O3} and \ce{TiO2}, thereby ensuring appropriate charge balance. (b) Equilibrium phase diagram based on Ref.~\onlinecite{kargin_phase_2015}. The compositional axis is rescaled from the original figure in Ref.~\onlinecite{kargin_phase_2015} to match the atomic fraction convention employed in this manuscript.}
        \label{fig:gibbs}
\end{figure*}

We also applied SARA-H to a second single composition system, the \ce{SnO_x} suboxide system, as a further benchmark to reproduce and validate the key findings of prior work by Sutherland~\cite{sutherland_autonomous_2023}. Using the same capped and uncapped sub-oxidized \ce{Sn-O} thin films, SARA-H’s autonomous learning framework recovered processing phase diagrams that closely match those previously obtained through exhaustive lg-LSA anneals and synchrotron-based experiments. The learned phase behavior reproduces the central conclusions of Sutherland, including the sequence of phase transformations from co-crystallized  \ce{SnO}/rutile \ce{SnO2} to full oxidation and eventual decomposition to metallic Sn, as well as the decisive role of atmospheric oxygen uptake during millisecond-scale anneals. In particular, uncapped films rapidly oxidize to \ce{SnO2}, whereas an \ce{Al2O3} capping layer effectively suppresses oxygen ingress and stabilizes \ce{SnO} across a broad processing window. The autonomously learned phase diagrams of the capped and uncapped samples are shown in Fig.~\ref{fig:SNOT} in the Supplemental Materials~\cite{SI}, while the associated learning curves shown in Fig.~\ref{fig:SNOT_Convergence} in the Supplemental Materials~\cite{SI} demonstrates rapid convergence, requiring fewer than 10 iterations for both the uncapped and capped films to reach $R_s^2=0.8$, indicating that the \ce{SnO_x} phase space is comparatively straightforward for SARA-H to learn.
Relative to the 105 experiments required in the original exhaustive study, this represents a resource reduction exceeding an order of magnitude. Full experimental details and analyses are provided in the Supplemental Materials~\cite{SI} in Sec.~\ref{sec:snot}.

\subsubsection{Active Learning on a Pseudo-Binary Composition Spread: Bi-Ti-O}
\label{sec:bitio}

A much more challenging autonomous experiment involves combining \ce{Bi_2O_3} with \ce{TiO2}.
To date, more than eight \ce{TiO2} polymorphs are known, with the most common crystalline phases comprising rutile ($\alpha$-\ce{TiO2}), anatase ($\beta$-\ce{TiO2}), and brookite ($\gamma$-\ce{TiO2}).
The latter two are metastable but are preferred in practical applications due to their higher photocatalytic activity~\cite{satoh_metastability_2013,cui_first-principles_2016,vu_anataserutile_2012,liu_photocatalytic_2012,di_paola_brookite_2013}.

The formation of various compounds in the ternary \ce{Bi2O3}-\ce{TiO2} system has been documented in the literature, garnering interest particularly due to the emergence of several material properties for diverse applications, including superior dielectric, piezoelectric, photorefractive, photocatalytic, or ferroelectric characteristics~\cite{zhou_synthesis_2006,newnham_structural_1971,shulman_microstructure_1996,wei_first-principles_2009,yao_photocatalytic_2004,hou_bismuth_2011,su_synthesis_2003}.
Kargin~\textit{et al.}~\cite{kargin_phase_2015} conducted a comprehensive study that summarized results from the literature in the form of an up-to-date equilibrium phase diagram for the \ce{Bi2O3}-\ce{TiO2} system, presented in Fig.~\ref{fig:gibbs}(b) for reference.
A full review of this material system is included in Sec.~\ref{sec:bitio_review} of the Supplemental Materials~\cite{SI}.
Based on the literature research, we set the goals of studying this system to be verifying the prediction of stabilization of anatase with Bi doping and studying the dynamics of binary oxide formation in this system.

We utilized SARA-H to investigate a thin film library of \ce{TiO2} and \ce{Bi2O3}, which was deposited by RF-reactive sputtering using elemental targets located at geometrically opposed positions.
This synthesized a compositional spread thin film spanning most of the \ce{Bi_2O_3}-\ce{TiO_2} pseudo-binary, as shown in Fig.~\ref{fig:BiTiO_XRF} in the Supplemental Materials~\cite{SI}.
To initiate the AL campaign, experimentally resolved crystal structures from the literature were gathered and 12 candidate phases were selected to seed both CrystalShift and the AL agents.
In addition to crystal structures of the unary and binary oxide of Bi-Ti-O systems \ce{SiBi2O5} was also included as \ce{Bi2O3} had been observed to react with the substrate to form a silicide upon melting.
With the exception of \ce{Bi36Ti32O}, all candidate phases lie on the tie line between \ce{Bi_2O_3} and \ce{TiO_2} in the Bi-Ti-O Gibbs triangle, as shown in Fig.~\ref{fig:gibbs}(a), ensuring the nominal oxidation state and associated charge balance.

\begin{figure}
    \centering
    \includegraphics[width=\linewidth]{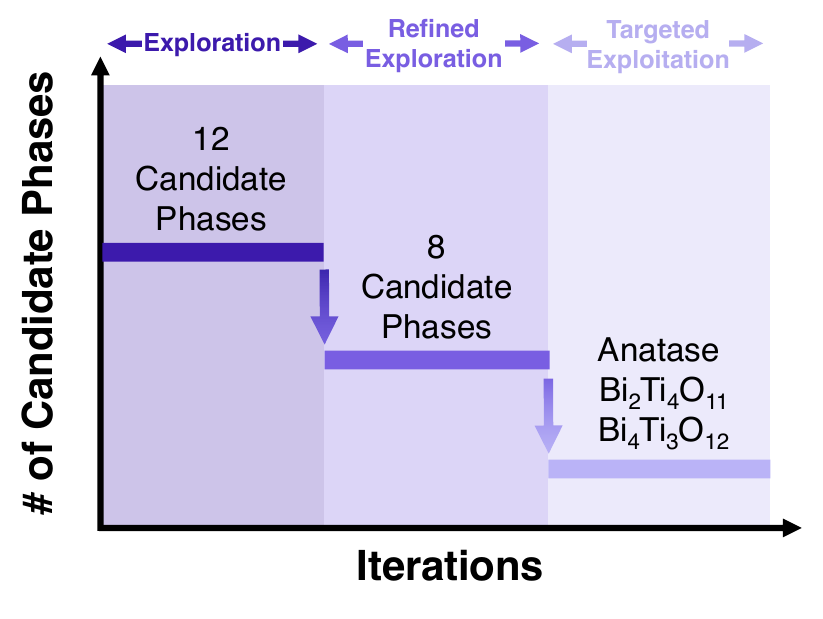}
    \caption{Stages of active learning for the Bi-Ti-O autonomous experimentation.}
    \label{fig:learning_stages}
\end{figure}

In contrast to unary systems, we progressively adjusted the candidate phase pool to guide the synthesis process for the Bi-Ti-O system, a procedure that roughly divides the AL campaign into three stages (Fig. \ref{fig:learning_stages}):

\begin{enumerate}
    \renewcommand{\labelenumi}{\roman{enumi})}
    \item Exploration (45 iterations): At the initial stage of the AL campaign, we deployed the cycle sampling acquisition function, considering all 12 candidate phases. The full set of candidate phases is listed in the Supplemental Materials~\cite{SI}.
    \item Refined Exploration (25 iterations): After initial exploration, the phase fields were visualized and inspected.
    Some phases were determined to be unsynthesizable using lg-LSA on our amorphous thin film precursor and were therefore removed to expedite the process.
    \item Targeted Exploitation (50 iterations): In the final stage, three phases were retained to focus on the objective that we had initially set.
\end{enumerate}

The phase-resolved learning curves of SARA-H's AL campaign are shown in Fig.~\ref{fig:BiTiO_Convergence} of the Supplemental Materials~\cite{SI}.
In contrast to the single-composition samples of \ce{Bi2O3} and \ce{SnO_x}, the Bi-Ti-O system is considerably more complex and challenging to learn, as evidenced by the much slower increase in the average $R_s^2$-score.
Approximately 80 iterations were needed for the average $R_s^2$ to exceed $R_s^2=0.8$.
Another subtle detail to note is the sudden increases in the $R_s^2$ scores for some of the individual phases.
These jumps may be associated with the cycle sampling strategy as shown in the benchmark section.

\begin{figure*}[ht!]
\centering
  \includegraphics[width=\linewidth]{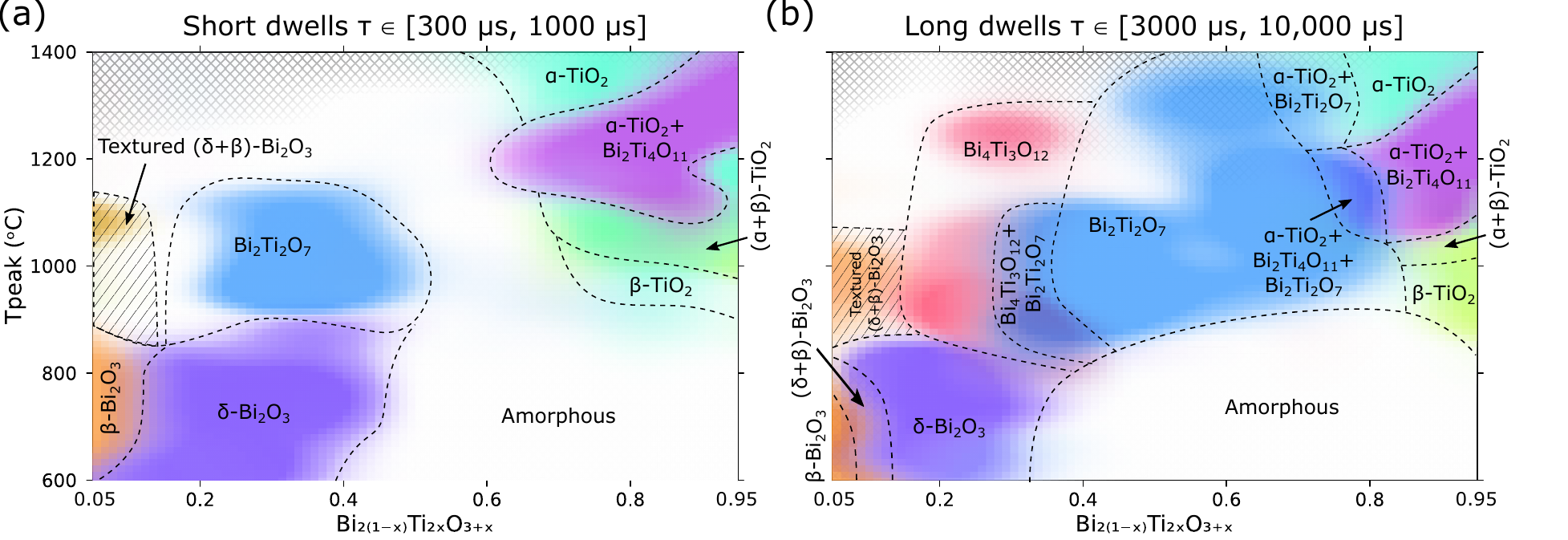}
  \caption{The experimentally determined processing phase diagrams of the Bi-Ti-O system after SARA-H's converged AL cycles with 120 iterations.
  Panel (a) displays the phase activations for the short dwell times of $\tau \in [300~\mu\text{s}, 1000~\mu\text{s}]$.
  Colors representing different phases are randomly selected from the HSV hue wheel.
  The intensity of each region is proportional to the corresponding phase activation.
  The amorphous domain is represented as the uncolored region.
  The dashed black lines denote the presumed phase boundaries and are merely intended as a guide for the eye. The cross hatch indicates the model's uncertainty and is more visible when the uncertainty is higher.
  Panel (b) shows the phase activations for the long dwell times of $\tau \in [3000~\mu\text{s}, 10,000~\mu\text{s}]$.}
  \label{fig:Bi-Ti-O}
\end{figure*}

Three-dimensional phase fields can be plotted as three-dimensional volume plots (Fig. \ref{fig:AL_phase_diagrams} in the Supplemental Materials~\cite{SI}), but it is challenging to extract quantitative information from 3D plots. 
The dwell dimension contributes the least variation in phase fraction, so to facilitate further analysis, we reduced the dimensionality by dividing the dwell domain into two halves and projecting out all $\tau$-dependence for each half.
The two resulting two-dimensional phase diagrams for short ($300 <\tau<  1000~\mu$s) and long dwell times ($3000 <\tau < 10,000~\mu$s) are shown as color-coded density plots in Fig.~\ref{fig:Bi-Ti-O}(a) and Fig.~\ref{fig:Bi-Ti-O}(b), respectively, where the expected phase activations are represented by colored regions.
The color saturation in each figure represents the probability densities, indicating the presence of the phases at the relevant processing peak temperatures and compositions. 

An immediately and clear feature of each phase map is the wide range of compositions that are associated with each phase, in marked contrast to the very narrow single-phase regions (i.e., line compounds) observed in equilibrium (Fig.~\ref{fig:gibbs}(b)).
This is readily understood by considering the short time during which the atoms have sufficient thermal energy to diffuse, which can both influence the crystal structure that forms and inhibits segregation into two-phase structures.

The results of short dwells illustrated in Fig.~\ref{fig:Bi-Ti-O}(a) 
reveal an extensive amorphous domain with two extrema: (i) for low temperatures and low Bi content the initially amorphous film does not overcome the energy barriers for nucleation and crystallization, and (ii) for high temperature and high Bi content the film undergoes rapid cooling from the melt, thus quenching in the amorphous structure. 
The existence of this second region is consistent with the trend of the equilibrium phase diagram (Fig.~\ref{fig:gibbs}(b)), which shows liquid phases in the high-temperature low Bi-content corner.

For intermediate values of Bi fraction and temperature ($0.5 < x < 0.7$ and $900 \ ^{\circ}\text{C} < T_{p} <1100 \ ^{\circ}\text{C}$) the films are also found to retain the as-deposited amorphous structure.
Careful manual examination of the XRD patterns in this region confirm this conclusion. 
The reasons for failure to crystallize films near equimolar Bi and Ti composition are not entirely clear; however, we expect that several factors contribute to the formation and shape of this amorphous region. 

First, the kinetics of crystallization play an important role: For \ce{Bi2O3} in Sec.~\ref{sec:bi2o3}, the crystallization onset at a dwell time of  $\tau=10,000~\mu$s is $T=280~^\circ$C, whereas the corresponding crystallization onset occurs at $T=480~^\circ$C for \ce{TiO2} (observed on a single composition \ce{TiO2} thin film sample).
It is thus not surprising that as-deposited  \ce{Bi2O3}-rich films exhibit lower activation energy for the nucleation of crystallites compared to \ce{TiO2}. 

Second, we anticipate a reduced driving force from the as-deposited amorphous state to the crystalline phases due to the increased compositional entropy when moving toward the region where both Bi and Ti are present in comparable concentrations. 
In the compositional range near $x=0.5$, the entropy of mixing is higher and the free energy difference to a low-entropy, ordered state is reduced.
Consequently, the dwell times of our anneals are too short compared to the required nucleation timescales.

As expected, for the most  Bi- and Ti-rich compositions, polymorphs of the unary oxides \ce{TiO2} and \ce{Bi2O3} are formed.
However, the $\delta$-\ce{Bi2O3} phase, which persists only in a narrow temperature range for pure \ce{Bi2O3} (see also Sec.~\ref{sec:bi2o3}),  dominates the phase diagram for $x > 0.1$.
This is not surprising since substitution by aliovalent ions is well known to stabilize the $\delta$-phase~\cite{gomez_stabilization_2014,durmus_electrical_2013,ling_review_1998,jung_structural_2002,struzik_defect_2012}.

Importantly, the pyrochlore phase \ce{Bi2Ti2O7} occurs over a wide composition range ($0.2 < x < 0.4$) on the Bi-rich side of the \ce{Bi4Ti3O12} phase region, despite having a lower Bi fraction.
Evidently in this composition range, the film first crystallizes into the Ti-substituted cubic $\delta$-\ce{Bi2O3} structure (an inference based on the assumption that the structure observed after quenching from peak temperatures below 900 $^{\circ}$C is formed during heating).
For regions heated to peak temperatures above 900 $^{\circ}$C,  the cubic structure transforms into pyrochlore \ce{Bi2Ti2O7} even though the thermodynamically stable phase is \ce{Bi4Ti3O12} and the stable composition range for \ce{Bi2Ti2O7} is very narrow (see Fig.~\ref{fig:gibbs}(b)).
We speculate that it is kinetically favorable to transform to the metastable pyrochlore structure rather than the Aurivillius phase structure (which has a lower free energy), as the latter requires more atomic hops to form.

For Ti-rich ($x> 0.6$) compositions, we observe an intricate progression of phases as the peak temperature is increased. 
For $900 \ ^{\circ}\text{C} < T_{p} < 1000 \ ^{\circ}\text{C}$, the amorphous film crystallizes into Bi-substituted anatase $\beta$-\ce{TiO2}, a metastable structure that is typically observed at low temperature.
At higher temperatures ($1000 \ ^{\circ}\text{C} < T_{p} <1100 \ ^{\circ}\text{C}$), a mixture of anatase and rutile structure is obtained, and at the very highest temperatures ($T_{p} \geq 1300 \degree$C) only rutile $\alpha$-\ce{TiO2} remains. 
Interestingly, at intermediate temperatures, $1100 \ ^{\circ}\text{C} < T_{p} <1300 \ ^{\circ}\text{C}$, $\beta$-\ce{TiO2} is no longer present but we observe $\alpha$-\ce{TiO2} coexisting with \ce{Bi2Ti4O11}, a phase expected from the equilibrium phase diagram in Fig.~\ref{fig:gibbs}(b), which nevertheless is not present at lower or higher temperatures.

In comparing the results of the longer dwell time to those of the short dwell times, it is readily apparent  that the high-temperature synthesis phase diagram (see Fig.~\ref{fig:Bi-Ti-O}(b)) exhibits more phases and larger crystalline phase fields.
In contrast to the case for short dwells, there are two disconnected amorphous regions:
the intermediate amorphous region observed for short dwells defined by $0.5 < x < 0.7$ and $900 \ ^{\circ}\text{C} < T_{p} <1100 \ ^{\circ}\text{C}$ in Fig.~\ref{fig:Bi-Ti-O}(a), now is part of an expanded phase field for \ce{Bi2Ti2O7}.
This is consistent with our interpretation that it is kinetics that inhibits crystallization. For longer dwells there is sufficient time for atoms to overcome activation barriers to nucleate crystallization.

The most striking feature of the long-dwell-time phase diagram is that the pyrochlore \ce{Bi2Ti2O7} completely dominates the  space within $0.4 < x< 0.8$ for  $T_{p} > 800  \ ^{\circ}$C.
This is unexpected since \ce{Bi2Ti2O7} appears as a line compound in the equilibrium phase diagram.
These results suggest that \ce{Bi2Ti2O7} nucleates first across a wide range of compositions due to a low activation barrier before it can transform into thermodynamically more stable phases at timescales much longer than those of our experiments.

For the  most Ti-rich compositions, we observe a progression similar to that seen at shorter dwells,  but the  stability and phase pure range of $\beta$-\ce{TiO2} is increased significantly.
Fig.~\ref{fig:bo_process} in the Supplemental Materials~\cite{SI} illustrates the way in which the relatively small phase volume identified as $\beta$-\ce{TiO2} evolves during the AL cycles.

The presence of substantial regions of pure $\beta$-\ce{TiO2} in our samples provides strong evidence that Bi acts as a dopant which inhibits the transformation of anatase to rutile in \ce{TiO2}.
This finding aligns well with the results of Hanaor~\textit{et al.}~\cite{hanaor_review_2011}, who predict that Bi will function as an inhibitor.
To the best of our knowledge, the present work is the first experimental evidence that doping \ce{TiO2} with Bi inhibits the anatase-to-rutile transformation.

Finally, we did not observe any indication of sillenite \ce{Bi12TiO20}, even after a thorough post-analysis of the XRD patterns from all anneals, neither at short nor at long dwell times.
This observation contrasts with the equilibrium phase diagram in Fig.~\ref{fig:gibbs}(b) and reports by Speranskyaya and Maier~\textit{et al}~\cite{speranskaya_system_1965,maier_notitle_1981}.
This finding is not entirely surprising, as the compositional domain where \ce{Bi12TiO20} should form is dominated by $\delta$-\ce{Bi2O3} which, as discussed, readily forms as the primary nucleating phase and which is capable of accommodating a significant level of Ti impurities due to its defective fluorite structure.

Overall, active learning of the synthesis phase diagram in this pseudobinary system has revealed a variety of unanticipated and complex phase evolution behaviors. Rather than following simple or monotonic trends, the observed phase behavior reflects the intricate interplay between synthesis conditions, kinetic pathways, and metastable phase formation.
Extending this approach to accumulate larger and more diverse datasets across a broader range of chemical systems will enable the identification of recurring patterns and correlations. Ultimately, such systematic, data-driven exploration is expected to facilitate the development of a more unified and predictive understanding of the principles governing metastable phase synthesis.

%% file: tex_files/conclusion.tex
We have significantly enhanced the capabilities of SARA, our AI-driven autonomous closed-loop materials discovery framework, by integrating on-the-fly XRD analysis and structural characterization with robotic materials synthesis using lg-LSA.
By leveraging an efficient and robust algorithm for probabilistic phase labeling, as implemented with NMF and CrystalShift, the enhanced SARA-H now can not only classify the crystal structures of synthesized metastable phases in a timescale commensurate with the pace of laser annealing and XRD experiments, but also autonomously conduct a targeted search for desired phases with high purity.
SARA-H's capabilities were highlighted by investigating three materials systems, namely \ce{Bi2O3}, \ce{SnO_x}, and \ce{Bi-Ti-O}. For \ce{Bi-Ti-O}, our targeted search notably resulted in identifying the processing space to form anatase, a polymorph of \ce{TiO2} that is particularly sought after due to its desirable photocatalytic properties and is stabilized by Bi doping.

To achieve efficient learning, we developed a novel sampling strategy based on the logEI acquisition function that balances the trade-off between the exploration and exploitation of aggregated information during AL experimentation, which we refer to as cycle sampling.
Extensive benchmarks demonstrate that cycle sampling significantly outperforms the baseline sampling methods (random and uncertainty sampling).
Most importantly, its target objective can be readily modified during an experimental campaign (i.e., on the fly) through the intervention of a human expert who can steer the search towards the targeted synthesis of specific phase(s).
Careful integration of a human-in-the-loop in AL designs not only accelerates the time to solution but has proven to be crucial in other AI domains where the well-orchestrated interplay between human and machine can exploit their individual advantages to ultimately optimize complex discoveries~\cite{mosqueira-rey_human---loop_2023,adams_human---loop_2023}.

While this work focuses on optimizing the targeted AL cycles specifically to maximize desired phase(s), the objective property can easily be substituted. 
For instance, we might aim to maximize unclassifiable diffraction patterns for the discovery of novel materials. 
Alternatively, we could incorporate additional characterization agents for property measurements, such as optical spectrometry to evaluate absorption for identifying synthesized materials suitable for solar energy applications, Raman spectroscopy that would reveal structural and chemical properties not exposed by XRD, or microscopic four-point probes for resistivity measurements.
A wide range of processing-structure-property relationships can be revealed by integrating these agents into SARA-H's AL cycles, and developing effective methods is a promising area for AI research.

The hierarchical multi-agent system described here integrates experimental materials processing and characterization with AI-based algorithms for reasoning and scientific discovery, including the representation, planning, optimization, and learning of materials knowledge. 
Our experimental methods are conducive to the formation of metastable materials which, broadly speaking, can exhibit unique and technologically valuable properties.
However, the synthesis pathways, stability ranges, and structure–property relationships of metastable phases remain poorly characterized, largely due to the difficulty of systematically exploring the vast and dynamic processing space in which they form.
By autonomously generating, characterizing, incorporating human-in-the-loop feedback, and iteratively refining experimental conditions, our system can produce rich, high-dimensional datasets that capture the emergence and evolution of metastable states.
With its capabilities, SARA-H acts as a powerful catalyst for the discovery of metastable materials, bridging the gap between theoretical potential and the rapid realization of next-generation technologies.

%% file: tex_files/methods.tex
\subsection{Thin Film Libraries}
Composition libraries were deposited on heavily \emph{p}-type doped (0.01~$\Omega$cm) silicon wafers (500$\mu$m thick).
Unless otherwise noted, all substrates had a 20~nm thermally grown oxide layer.
Gold alignment features were first formed on wafers using standard lithography.
Thin films were then deposited on the wafers using radio frequency (RF) magnetron sputtering using 2-inch diameter targets for the corresponding material system.
The atmosphere and detailed sputtering conditions for each film are as follows:

\begin{description}
    \item[Bi-O system] RF sputtering from a \ce{Bi} target in an atmosphere of 8~mTorr Ar and 2~mTorr \ce{O2} to deposit the \ce{Bi2O3} sample on 200~nm thermally oxidized \ce{SiO_2} wafers.
    Wafers were rotated while sputtering at an RF power of 20~W to produce a 170~nm thick film with $<10\%$ thickness variation across the wafer.
    \item[Sn-O system] 
    Thin films of SnO$_{x}$ were sputtered from a metallic Sn target in an \ce{O2} and \ce{Ar} gas mixture of 6\% \ce{O2} at a total pressure of 5~mTorr in an AJA International deposition system.
    Samples were rotated to form $\approx100$~nm thick, predominantly amorphous, films.
    Half of the wafer was capped with a $\approx$15~nm thin film of \ce{Al2O3} sputtered from an \ce{Al2O3} target to reduce \ce{O2} exposure of the film during laser annealings.
    \item[Bi-Ti-O system] The (Ti,Bi)\ce{O_x} composition spread films were sputtered using 2-inch diameter metal targets in two Angstrom~\cite{angstromengineering} sputter guns arranged in the 90$^\circ$ off-axis configuration to minimize bombardment of the growing film by energetic O anions in the plasma~\cite{dover_codeposited_2004}.
    The Ti target ($>99.9\%$ pure) was RF powered at 125~W and the Bi target ($>99.99\%$ pure) at 15~W RF, while the Si substrate was RF powered at 6~W, resulting in a DC bias of -40~V to help densify the as-grown film.
    The sputtering ambient was $20\%$ \ce{O2}, $80\%$ Ar supplied at an overall flow rate of 30~SCCM at 30~mTorr.
    The substrate was unheated during sputtering and as-deposited films were amorphous.
    The cation composition range was roughly $0.1 < \text{Bi/(Bi+Ti)} < 0.9$ (detailed compositional characterization post-deposition is reported in Sec.~\ref{sec:comp_characterization} together with the XRF measurements in Sec.~\ref{sec:xrf} of the Supplemental Materials~\cite{SI}).
    The film was deposited for 60 minutes yielding thicknesses ranging from 200 to 300~nm over the composition spread.
    As with Ti-O, no thickness dependent variations in phase formation are observed within this range.
\end{description}

\subsection{Compositional Characterization for Bi-Ti-O\label{sec:comp_characterization}}
The composition of binary spreads was evaluated using X-ray fluorescence (XRF) using an EDAX Orbis Micro-XRF system with a 2~mm X-ray beam diameter.
The Bi~K and Ti~K XRF peak intensities were converted to normalized cation compositions using sensitivity factors calibrated by commercial XRF standards (Micromatter\texttrademark{}). 
The measured composition of the film is presented as a colored scatter plot in Fig.~\ref{fig:BiTiO_XRF} of the Supplemental Materials~\cite{SI}.
Since the density of the XRF data was lower than that of the lg-LSA stripes on the wafer, the data were interpolated using a smooth GP model with prior length scales of $\sigma_{x,y}=2$~mm and an RBF kernel.

\subsection{Lateral Gradient Laser Spike Annealing\label{sec:lglsa}}
The lateral gradient laser spike annealing (lg-LSA) was based on the design introduced by Bell~\textit{et al} \cite{bell2016lateral}.
A 120~W \ce{CO2} laser ($\lambda=10.6~\mu$m) was passed through an attenuator to control the incident power on the wafer, calibrated with a power meter positioned at the wafer plane.
The incidence angle of the laser with respect to the wafer normal was $73^\circ$, close to Brewster's angle to minimize reflected power.
Due to this shallow incidence angle, the temperature profile exhibited a two-sided Lorentzian intensity distribution on the wafer with the laser intensity displaying different widths on either side of the peak power.

To secure the wafer in place and ensure consistent thermal contact, a porous aluminum vacuum wafer chuck was used.
For each anneal, the stage was first accelerated to the desired velocity corresponding to the desired dwell time.
Once the velocity was reached, a shutter opened to  anneal the sample.
After the anneal stripe reached the designed length (5 mm), the shutter was closed and the stage was decelerated to a complete stop.

To quantify the lateral temperature distribution and provide the active learning agent with dense phase-temperature information, thermoreflectance was used to calibrate the temperature field of lg-LSA.
This method leverages the known linear relationship between silicon’s reflectance and temperature to enable accurate thermal mapping.
Following the process described elsewhere~\cite{chang2025enabling}, the thermoreflectance coefficient, and thus the absolute temperature, was calibrated using the known melting points of gold and silicon as references.
Surface functions were then fitted to the parameterized temperature distribution in the processing space and used to reconstruct the temperature distribution under arbitrary experimental conditions within the range of interest.

\subsection{Temperature Calibration of lg-LSA\label{sec:calibration}}
To extract the dense temperature–phase information from lg-LSA stripes, we established an accurate calibration of the full temperature profile generated during annealing. For a given laser power $P$ and scan velocity $v$, the goal is to determine the peak anneal temperature distribution $T_p(v, P, x)$ at every position $x$ \textit{across} the stripe.
We calibrated this temperature profile at a fixed downstream position $y$ \textit{along} the stripe, namely at the $y=4$~mm mark of an anneal stripe, chosen to ensure quasi-equilibrium laser conditions while avoiding artifacts associated with shutter actuation.

Accurate temperature calibrations were obtained by employing a thermoreflectance approach, which leverages the linear dependence of silicon's optical reflectivity on absolute temperature change~\cite{abel_noncontact_2006}.
By monitoring relative reflectance changes during laser scans with a camera and associating them to a reference reflectance at a known temperature, we obtained a robust, spatially resolved measure of the peak thermal response under each annealing condition.

A comprehensive calibration dataset was assembled by repeating the reflectance measurements over a wide range of laser powers and scan velocities and was subsequently used to fit a parametrized temperature model.
Instead of modeling the peak anneal temperature distribution $T_p(v, P, x)$ as a single function, we constructed empirical functional relationships that describe the peak temperature at the center of the stripe ($T_c(v, P)$) and the lateral extent along $x$ of the temperature distribution, modeled as a two-sided Lorentzian curve with widths $R(v, P)$ and $L(v, P)$ for the right and left sides, respectively.
To convert relative temperatures from reflectance measurements into an absolute temperature scale, the model was calibrated by experimentally determined melting conditions of silicon and gold, whose melting points are well established.
Once calibrated, the center peak temperature and the width distributions were converted from scan velocities $v$ to dwell times $\tau = \frac{\text{FWHM}}{v}$, where FWHM denotes the full width at half maximum of the laser power profile along the scanning direction $y$. 
The combined hypersurfaces of  $T_c(\tau, P)$, $R(\tau, P)$, $L(\tau, P)$ enable prediction of complete temperature profiles as a function of laser power and dwell time with an accuracy of approximately~$\pm25~^\circ$C near the center of a stripe.
A detailed description of thermoreflectance modeling, fitting procedures, and calibration strategy was described elsewhere~\cite{chang2025enabling}.

\subsection{High-Throughput Wide Angle X-Ray Diffraction}
The lg-LSA anneal stripes are nominally $\approx 1.5$~mm wide, resulting in temperature gradients of $\approx 1-5~^\circ\text{C}/\mu\text{m}$, depending on laser anneal conditions.
To sample this temperature distribution with sufficient spatial resolution and in a high-throughput manner, we utilized wide-angle X-ray scattering at the Cornell High Energy Synchrotron Source (CHESS) ID3B beamline~\cite{smieska_functional_2023}, employing a 9.7~keV X-ray beam focused by a compound reflective lens to a spot size of $20\times60~\mu$m upon striking the thin film sample at a $3^\circ$ incident angle. 
The diffracted X-ray signals were collected by a Dectris Eiger 1M detector.

To collect spatially resolved X-ray diffraction across an annealed stripe, the detector continuously gathered 2D X-ray diffraction data at 50~ms intervals while the lg-LSA precision stage stepped horizontally relative to the X-ray beam location.
The stepping velocity was set to achieve a spatial resolution of $10~\mu$m, which corresponds to a temperature resolution better than $\Delta T \approx 50~^\circ$C.
A total of 151 two-dimensional X-ray diffraction images were collected for each anneal stripe, covering a distance of 1.5~mm.
An open-source azimuthal integration package, PyFAI~\cite{ashiotis2015fast}, was used to integrate the 151 2D X-ray diffraction images in parallel into 1D XRD patterns on-the-fly, utilizing the high-performance computing facility at CHESS.
It is critical to carefully characterize the scattering geometry since small uncompensated variations can lead to large errors in the scattering vector due to the low angle of incidence of the X-ray beam.
A representative heat map of a scanned XRD map is shown in the lower portion of the second panel of Fig.~\ref{fig:workflow}.

\subsection{Automated X-ray Diffraction Analysis and Phase Labeling}
In the first step of processing the acquired, integrated XRD patterns of an annealed stripe, nonnegative matrix factorization (NMF) was performed to reduce the dimensionality of the data and the subsequent demand on computational resources.
The NMF can be accomplished by solving the following approximation problem
\begin{equation}
\label{equ:nmf}
   \begin{split}
        \mathbf{X} &\approx \mathbf{WH}\\
        \text{s.t.} \mathbf{W}&\geq0 \text{, } \mathbf{H}\geq0
    \end{split} 
\end{equation}
where $\mathbf{X}\in\mathbb{R}^{n\times m}$ is the data matrix, in which each column represents an XRD pattern, and the inequalities apply element-wise.
For this reason, $\mathbf{W}\in\mathbb{R}^{n\times k}$ is a matrix whose columns can be interpreted as prototypical non-negative XRD patterns that are added together in different proportions using the non-negative coefficients in  $\mathbf{H}\in\mathbb{R}^{k\times m}$.
Due to the matrices' functions, we refer to $\mathbf{W}$ as the basis matrix and $\mathbf{H}$ as the activation matrix.
The integer $k$ is a user-defined parameter that represents the rank of the data matrix.
In the present implementation, we applied an NMF that is based on an extreme ray finding algorithm~\cite{kumar2013fast}.
Notably, the objective of this algorithm is to select representative columns in  $\mathbf{X}$ to construct a basis set that minimizes Eq.~\eqref{equ:nmf}, rather than to separate the phases in a mixture.
We addressed this separation problem with other methods below.
This selection process avoids the formation of non-physical basis XRD patterns.
In AL our campaigns, we used $k=4$ and attributed the basis that has the highest activation value at the first column of the XRD data to be the signal from the as-deposited film and discarded it from subsequent data processing.

To reduce computational cost and avoid labeling non-crystalline signals, we fit each basis with a smooth background model using kernel ridge regression (RBF kernel).
Bases with fitting errors below a threshold were classified as amorphous and discarded.
The threshold was set conservatively to ensure that crystalline signals are not inadvertently removed.

Next, CrystalShift, a rapid probabilistic multiphase labeling algorithm was employed to label phases and estimate the phase fractions associated with each basis vector in $\mathbf{W}$~\cite{chang2023probabilistic}.
CrystalShift effectively integrates a best-first tree search, rapid lattice refinement, and Bayesian model comparison to provide probability estimates for potential phase combination assignments.

Here, we briefly describe the high-level concept and process of the algorithm with further details available in the original papers.
CrystalShift uses a best-first tree search scheme to explore the combinatorial space of phases.
In the tree, each node represents a possible phase combination.
At each level of the tree, CrystalShift performs a pseudo-refinement process on all nodes at that level.
For each factorized XRD pattern basis vector $\mathbf{W}_i$, the pseudo-refinement algorithm minimizes the difference between the simulated diffraction pattern of the phase combination and the given pattern by modifying the crystal system preserving parameters $\bm{\theta}$ of a given candidate phase set
\begin{equation}
    \label{eq:opt}
    \bm{\theta}^*=\arg\min_{\bm\theta} ||f(\bm{\theta})-\mathbf{W}_i||^2_2
\end{equation}
where $f$ is the function that computes the XRD pattern from the parameters $\bm\theta$.
Then, the nodes with the $k_{\mathrm{exp}}$ lowest residuals  are expanded, which means adding an additional phase that was not originally present in the node to the next level.
In this work, $k_{\mathrm{exp}}=3$ was used.
Starting with an empty root node at level 0, CrystalShift repeats the refine-and-expand loop until the desired level is reached, as defined by the maximum number of allowed coexisting phases specified by the user. 

After the tree search is completed, the likelihood of each node is calculated by marginalizing out the parameters~$\boldsymbol{\theta}$ of the model.
Since this  process is mathematically intractable, the Laplace approximation is used to perform the marginalization.
From Eq.~\eqref{eq:opt}, the log-likelihood of a set of candidate phases $M_i$ can be written as
\begin{equation}
    \log p(\mathbf{t}|M_i, \bm{\theta}^*_i) = -L_i.
\end{equation}
where $L_i=||f(\bm{\theta^*}_i)-\mathbf{W}_i||_2^2$ is the residual of Eq.~\eqref{eq:opt}.
Using Laplace approximation to marginalize out $\boldsymbol{\theta}$ gives
\begin{equation}
    \log p(\mathbf{t}|M_i) \approx -L_i - \frac{\log|H_{\bm{\theta}^*_i}(L_i)|}{2} + d_i\frac{\log(2\pi)}{2}
\end{equation}
where $H_\mathbf{{\bm{\theta}}^*_i}(L_i)$ is the Hessian of the loss with respect to the variables at the optimum and $d_i$ is the dimensionality.
To ensure the probabilistic labeling was well-calibrated, a temperature scaling was used~\cite{guo2017calibration}, which works by multiplying the resulting log-likelihood by a temperature parameter $\beta$.
The temperature parameter was determined by performing thousands of labelings on a calibration dataset containing 10,000 synthetic XRD patterns  that approximately span the same q-range as our experiment.
The XRD calibration dataset was generated using candidate phases used in each AL trial and each pattern contained up to two phases with symmetry-allowed lattice strains $\epsilon \in [-0.05, 0.05]$, phase fractions $p_i \in [0.25, 0.75]$, and a random peak width factor $\sigma \in [0.1, 0.3]$.
The temperature parameter was selected such that it minimizes the mean absolute error between the calibration curve and the ideal diagonal line in the reliability diagram~\cite{hamill1997reliability}.
Finally, the scaled log-likelihood passes through a softmax function to produce a probability distribution over all possible phase combinations given the experimental XRD pattern.
The posterior probability for each set of candidate phases $M_i$ is then given by
\begin{equation} 
    p(M_i | \mathbf{t}) = \frac{\exp \left( \beta \log p(\mathbf{t}|M_i) \right) p(M_i)} {\sum_{j=1}^N \exp \left( \beta \log p(\mathbf{t}|M_j) \right) p(M_i)},
\end{equation}
where usually the prior over phase combinations is uniform, i.e. $p(M_i) = 1 / N$.

To enhance SARA-H's robustness against experimental uncertainty and phase labeling errors, we provide expected phase activations that incorporate the probability estimates generated by CrystalShift, rather than directly reporting the most probable phase fraction to the AL agent.
A combinatorial sum over the top candidates of each basis is performed to combine the results.
Here, we consider only the top 5 candidates for each of the 3 bases and produce 125 potential phase models for the stripe.
The final phase probabilities produced by this model are then the product of all probability estimates of each phase model $M_i$, re-normalized so that the probabilities of all 125 models sum to 1.

Finally, we calculate the expected phase activations of all phases by
\begin{equation}
    \mathbb{E}[a_{j}^{(i)}]=\sum_{k=1}^K H_{kj} \mathbf{p}^T  \mathbf{a}^{(i)}
\end{equation}
where $a_{j}^{(i)}$ represents the activation of phase $i$ at position index $j$, $H_{kl}$ is the activation matrix element at the $k$-th row and $l$-th column, $ \mathbf{p}$ is the probability vector of the 125 models, and $\mathbf{a}^{(i)}$ is the collected vector of the activation of phase $i$ in each of the 125 models.
The expected activation of phase $i$ at location index $j$, $\mathbb{E}[a_{j}^{(i)}]$, is then mapped to the expected activation of phase $i$ at temperature $T$, $\mathbb{E}[a_{j}^{(T)}]$, using the temperature profile introduced in Sec.~\ref{sec:calibration}.
The activation is smoothed using a Gaussian process (GP) with a square exponential kernel.
To avoid secondary effects that may occur at the edges of the XRD map, i.e., the tails of the temperature profile, and to ensure that the temperature gradient is not too steep and, therefore, that the thermal resolution was sufficiently high, we sampled 16 points between the peak temperature and 200~$^\circ$C below the peak temperature with equal spacing.

\subsection{Active Learning for Targeted Phase Mapping}
\label{sec:activelearning}
\subsubsection{Gaussian Process Models\label{sec:almodel}}
The goal of active learning is to reason with the currently available expected phase activation distribution to suggest the next experimental condition that would yield the largest expected improvement of the phase diagram.
We utilized an AL scheme based on Bayesian optimization, which includes a surrogate model and an acquisition function for experimental decision making.
In our cyclic AL setup, a Gaussian process (GP) regression model was chosen as the surrogate model because of its reliable uncertainty estimates and its capability to incorporate prior domain knowledge through kernel function constructions.

A GP is a distribution over functions, and its behavior is entirely determined by its mean $\mu(\cdot)$ and covariance function $K(\cdot, \cdot)$, commonly referred to as the kernel function.
Given a set of currently available input data in matrix form $ \mathbf{X}$, a GP can be expressed as~\cite{williams2006gaussian}
\begin{equation}
    f_{\mathcal{GP}}( \mathbf{X}) \sim \mathcal{N}(\mu( \mathbf{X}), K( \mathbf{X},  \mathbf{X}))
\end{equation}
where $K( \mathbf{X},  \mathbf{X}')$ is the covariance matrix, with its ($i$-th,$j$-th) entry equal to $K( \mathbf{X}_i,  \mathbf{X}_j)$.
When performing GP inference on a new point $ \mathbf{x}^*$, the GP posterior distribution is generated by conditioning on the joint distribution of the GP prior to obtain the posterior.
This process results in another multivariate Gaussian distribution $f_p$,
\begin{gather}
    f_p = p(f| \mathbf{X},  \mathbf{x^*},  \mathbf{y}, f_{\mathcal{GP}})=\mathcal{N}(\mu_p( \mathbf{x}^*), \Sigma_p( \mathbf{x}^*,  \mathbf{x}^{*'})) \\
    \mu_p( \mathbf{x}^*) = K( \mathbf{x}^*,  \mathbf{x})K( \mathbf{X},  \mathbf{X})^{-1}( \mathbf{y}-\mu({ \mathbf{X}}))\\
    \Sigma_p( \mathbf{x}^*,  \mathbf{x}^{*'}) = K( \mathbf{x}^*,  \mathbf{x}^{*'})-K( \mathbf{x}^*,  \mathbf{X})K( \mathbf{X},  \mathbf{X})^{-1}K( \mathbf{X},  \mathbf{x}^*)
\end{gather}
The kernel used in this work was a Matérn-5/2 kernel with length scales of $\sigma_{T_p}=50$, $\sigma_{\tau_{\text{log}}} = 0.3$, and $\sigma_c = 0.15$ for the dimensions of peak temperature $T_p$, the decadic logarithm of the dwell time $\tau$, and the composition axis $c$ if applicable, respectively.
In principle, the length scales of the Matérn kernel can be optimized by maximizing the marginal likelihood~\cite{williams2006gaussian}.
However, our selection of the aforementioned length scales was based on benchmarks utilizing a range of fixed values, from which we chose the best-performing combination for the kernel~\cite{ament_autonomous_2021}.

\subsubsection{Uncertainties\label{sec:uncertainty}}
It is critical to consider all sources of uncertainty when designing nested learning strategies that rely on experimental, noisy data.
As outlined in our previous work~\cite{ament_autonomous_2021}, we accounted for errors in the measurements (i.e., model outputs from the XRD characterization) as well as intrinsic uncertainties in the temperature profile.
We approximated the posterior inference with input noise by propagating the input uncertainty using a linear approximation of the standard posterior mean, following the approach of McHutchon and Rasmussen \cite{mchutchon2011input}.
In this model, given the regular posterior mean $\mu_p(x)$, its input-noise-corrected version can be computed by updating
\begin{equation}
\label{eq:nigp}
 K \leftarrow K + \text{diag}(\sigma_x(\mathbf{X}) \odot \partial_x \mu_p(\mathbf{X}))^2 
\end{equation}

Since the phase activation is calculated as a function of position, we need to convert the stripe-specific activation function $a_{k,T_c, \tau}(x)$ of phase $k$ to the temperature domain using the temperature profile $T_{T_c, \tau}$, resulting in $a_{k,T_c, \tau}(T)$.
Given that there exist experimental uncertainties in position $x$, and consequently in $T$, it is essential to quantify the uncertainty arising from these input errors and propagate them to the phase activation model. 

The intrinsic experimental uncertainties resulting from the temperature profile $T_{T_p, \tau}(x)$ of the laser exhibit a variance around the true value as a function of position 
\begin{equation}
\label{eq:temperature_uncertainty}
       \sigma_T^2(x) = \sigma^2_{T_c} \left(
        \frac{T_c}{1400}\right)^2 
        + \sigma_x^2 \ \left(\frac{\partial T_{T_c, \tau}(x)}{\partial x}\right)^2
\end{equation}
where $\sigma_{T_c}$ denotes the standard error in the peak temperature and $\sigma_x$ indicates the standard error in the position.
The first term quantifies the error at the peak temperature, which is largest at high temperatures ($1400^{\circ}$C) and decreases linearly with $T$, due to the uncertainty in the conversion factor $\kappa$ in the thermoreflectance measurement.
The second term quantifies the uncertainties in the temperature profile, which arise primarily due to the limited spatial resolution.
The form of the uncertainty term is derived using standard error propagation techniques~\cite{tellinghuisen2001statistical}.
We estimate $\sigma_{T_c} = 25^{\circ} \text{C}$ and $\sigma_x = 50 \mu\text{m}$ based on the mean square error during the fitting of the temperature profile, the convolution of the X-ray intensity distribution, and the step size.
We then use the uncertainties from Eq.~\eqref{eq:temperature_uncertainty} with Eq.~\eqref{eq:nigp} to update the uncertainties in phase activations.

\subsubsection{Acquisition Function}
We selected a recently developed flavor of the expected improvement (EI) called LogEI as our core acquisition function~\cite{ament2023unexpected}.
LogEI has been shown to be superior to conventional EI because it addresses issues arising from vanishingly small gradients by providing a more numerically robust way of computing the acquisition function. This method has already been applied successfully for the optimization of sustainable concrete~\cite{ament2023sustainableconcretebayesianoptimization}, a mixture composition optimization problem.

To briefly summarize: given a GP posterior mean $\mu$ and variance $\sigma$ as well as the currently maximal phase activation at $y^*$, the logEI acquisition function evaluated on an experimental condition vector $\mathbf{x}$ is given by
\begin{equation}
    \text{LogEI}_{y^*}(\mathbf{x}, y^*)=\log \left( h\left[\frac{\mu(\mathbf{x})-y^*}{\sigma(\mathbf{x})}\right] \right) + \log\left(\sigma(\mathbf{x})\right)
\end{equation}
where $h(x)=\phi(x)+\sigma(x)\Phi(x)$ and $\phi(x), \Phi(x)$ are the probability density function and cumulative distribution function of a normal distribution, respectively.
However, the first term, if implemented naively, can still encounter numerical underflow issues.
To solve this problem, the function composition $\log \circ h$ can be calculated in a numerically stable manner by
\begin{widetext}
\begin{equation}
    \log \circ h (z) =
    \begin{cases}
        \log(\phi(z)+\sigma(z)\Phi(z)) & z>-1 \\
        -z^2/2 -\log 2\pi/2 + \texttt{log1mexp}(\texttt{logerfcx}(z/\sqrt{2})|z|+\log(2\pi)/2 )) & z <=-1
    \end{cases}
\end{equation}
\end{widetext}
where \texttt{log1mexp} and \texttt{logerfcx} are numerically stable implementations of $\log(1-\exp(z))$ and $\log(\exp(z^2)\text{erfc}(z))$, respectively. With this implementation, we can accurately compute $\log \circ h (z)$ with $z \in [10^{-100}, 10^{100}]$.
We optimized LogEI with the BFGS algorithm~\cite{liu_limited_1989,press_numerical_2007} to find the optimal $\mathbf{x}$.
The best-found $\mathbf{x}$ then determined the composition and processing conditions for the next lg-LSA experiment.

\subsection{Synthetic Test for Different Sampling Methods\label{sec:synthetictests_methods}}
\subsubsection{Synthetic Model\label{sec:synthmodel}}
We benchmarked various sampling strategies using a synthetic data model that is based on experimentally obtained and hand-labeled Bi-Ti-O phase data, aiming to replicate realistic performance conditions as accurately as possible.
To create the benchmark system, we incorporated eight distinct phases, each modeled by a GP with a Matérn 5/2 kernel based on the data.
Each phase covers different regions and has a different center, representing the condition of maximal phase activation in the parameter space, defined by the temperature $T$, the decadic logarithm of the dwell time $\tau$, and one compositional dimension $c$.
To speed up the testing process, we first evaluated all GPs on a fixed condition grid and interpolated linearly using the three nearest points on the condition grid during tests.

We included four distinct sampling methods in our benchmarks:
\begin{itemize}
    \item Random sampling: The experimental conditions are generated randomly in each iteration of the AL process. 
    \item Uncertainty sampling: The experimental conditions are selected where the average uncertainty of the model over all candidate phases is maximal in each iteration of the AL process. 
    \item Target sampling: A single simulated pattern of the phase(s) present at the sampled experimental condition is used as the input to CrystalShift. The expected activation of one phase is then used as the objective to maximize in every AL iteration.
    \item Cycle sampling: A single simulated pattern of the phase(s) present at the sampled experimental condition is used as the input to CrystalShift. At each iteration of the AL process, we cycle through the expected activation of different phases in our candidate pool, serving as the objective for active learning. That is, assuming there are three candidate phases $\alpha$, $\beta$, and $\gamma$, the cycle sampling strategy will optimize the expected activation of phases $\alpha, \beta, \gamma, \alpha, \beta, \dots$ in the AL iterations $0, 1, 2, 3, 4, \dots$.
\end{itemize}

In addition to the sampling strategies mentioned above that consider only data associated with the peak temperature $T_c$ at the center of an anneal, we also introduced an extension called ``stripe sampling'': the model utilizes the results from $T_c$ to $T_c-200~^\circ$C when probing a condition of $(T_c, \tau,c)$ and the acquisition function  sums the expected improvement over the same temperature range in order to make use of the temperature gradient for improved decision making.
The stripe extension is applied to enhance each of the aforementioned AL methods.

For the benchmark runs, we included twelve phases in the candidate pool.
These candidate phases are identical to those used in our real experiments described in Sec.~\ref{sec:bitio} (also listed in the Supplemental Materials~\cite{SI}).
The Bayesian optimization setup for the AL model was identical to the one reported in Sec.~\ref{sec:almodel}, utilizing a Matérn-5/2 kernel with length scales of $\sigma_{T_c}=50$, $\sigma_{\tau_{\text{log}}} = 0.3$, and $\sigma_c = 0.15$ for the dimensions of peak temperature $T_c$, the decadic logarithm of the dwell time $\tau$, and the composition axis $c$, respectively.
Each approach was repeated 500 times to gather statistically meaningful averages.
The stripe sampling variants were repeated 100 times due to their longer run times.

\subsubsection{Performance Metrics\label{sec:metrics}}

We evaluated the effectiveness of the active learning strategies using various performance metrics, which are formally defined in this section to provide a clear basis for our benchmarking results.
First, the regret $r_k$ for each phase $k$ is defined as $r_k=|a_k(\hat{\mathbf{x}})-a_k(\mathbf{x}^*)|$, where $a_k$ represents the regression model and maps the parameters within the space of $\{\mathbf{c}, \tau, T \}$ to its predicted activation $a_k({\mathbf{x}})$.
This measure represents the relative difference between the maximum phase activation by the AL model, $a_k(\mathbf{x}^*)$, and its true maximum phase activation, $a_k(\hat{\mathbf{x}})$.

Next, we introduce the interquantile range (IQR) of regret, which is defined as the difference between the \nth{75} and \nth{25} percentiles of the regret distribution in the statistical ensemble. 

The enhancement factor (EF) at iteration $i$ for phase $k$ is defined by
\begin{equation}
    \text{EF}_{\text{i}}^k = \frac{a_{\text{AL},k}(\mathbf{x}_{1:i}^*)}{a_{\text{random},k}(\mathbf{x}_{1:i}^*)}
\end{equation}
where $a_{\text{AL},k}(\mathbf{x}_{1:i}^*)$ is the maximum phase activation observed before iteration $i$ of phase $k$.
The EF curve illustrates the ratios between the objective values in the AL runs and the random sampling runs at various stages, indicating how much improvement the AL techniques provide in terms of objective values.

The acceleration factor (AF) is defined as
\begin{equation}
    \text{AF}^k(a) = \frac{\text{i}_{\text{random}, k}(a)}{\text{i}_{\text{AL},k}(a)}
\end{equation}
where $\text{i}_{\text{s},k}(m)$ is given by $\arg \min_i a_{\text{s},k}(\mathbf{x}_{1:i}^*) \geq m$ for phase $k$ with s being the sampling strategy.
In other words, the AF measures the speed-up achieved by the AL strategy relative to random sampling in reaching the same objective value.

The coefficient of determination $R_k^2$ for a phase $k$ is evaluated on the full model, including all of the $n$ acquired data points $i$, and is defined as 
\begin{equation}
    R_k^2 = 1 - \frac{\sum\limits_{i=1}^n(a_k(\mathbf{x_i})-y_i^k)^2}{\sum\limits_{i=1}^n(\mu(\mathbf{y}^k)-y_i^k)^2}
\end{equation}
where $y_i^k$ represents the ground truth activation of phase $k$ at coordinate $\mathbf{x_i}$, and $\mu(\mathbf{y}^k)$ denotes the mean of the data $\mathbf{y}^k$. 
Furthermore, the mean coefficient of determination $R^2=\bar{R}_k^2$ is the average over $R_k^2$ of all phases $k$.
However, since we are dealing with heteroscedastic data, we generalized $R_k^2$ by weighting over the standard deviation $\sigma_i$ at each data point $i$, the concept of which is based on the log-likelihood of the heteroscedastic normal errors.
This approach leads to
\begin{equation}
    R_{S_k}^2 = 1 - \frac{\sum\limits_{i=1}^n(a_k(\mathbf{x_i})-y_i^k)^2/\sigma_i^2}{\sum\limits_{i=1}^n(\mu(\mathbf{y}^k)-y_i^k)^2/\sigma_i^2}
\end{equation}
and its associated mean coefficient of determination $R^2_S=\bar{R}_{S_k}^2$ as the average over all phases $k$.

%% file: tex_files/acknowledgements.tex
The authors acknowledge the Air Force Office of Scientific Research for its support under award FA9550-18-1-0136.
This work is also supported by the Air Force Office of Scientific Research (AFOSR), Defense University Research Instrumentation Program (DURIP) FA9550-23-1-0569 and the Schmidt Sciences AI2050 Senior Fellowship.
This work is based on research conducted at the Materials Solutions Network at CHESS (MSN-C) which is supported by the Air Force Research Laboratory under award FA8650-19-2-5220, and the National Science Foundation Expeditions under award CCF-1522054.
This work was partially performed at the Cornell NanoScale Facility, a member of the National Nanotechnology Coordinated Infrastructure (NNCI), which is supported by the National Science Foundation (Grant NNCI-2025233).
This research was conducted with support from the Cornell University Center for Advanced Computing.
The authors would also like to acknowledge the technical assistance provided by staff at CHESS.

%% file: tex_files/contributions.tex
MCC and MA took the lead in writing the manuscript.
MCC, MA, and SA developed and implemented the algorithms to execute SARA-H.
DRS, RBvD, and KRG fabricated the thin film samples. 
LZ and JMG provided compositional characterization of the thin films.  
LMS and ARW offered methodological and technical support for the XRD measurements. 
RBvD, CPG, MOT, and JMG supervised the research.
MCC and MA contributed equally to this work. 
All authors provided critical feedback and helped shape the research, analysis, and manuscript.

%% file: tex_files/SI.tex
\subsection{Supplemental Benchmark Figures}
This section contains additional information to support the performance benchmarks discussed in the main manuscript.

\begin{figure*}[h!]
    \includegraphics[width=0.6\linewidth]{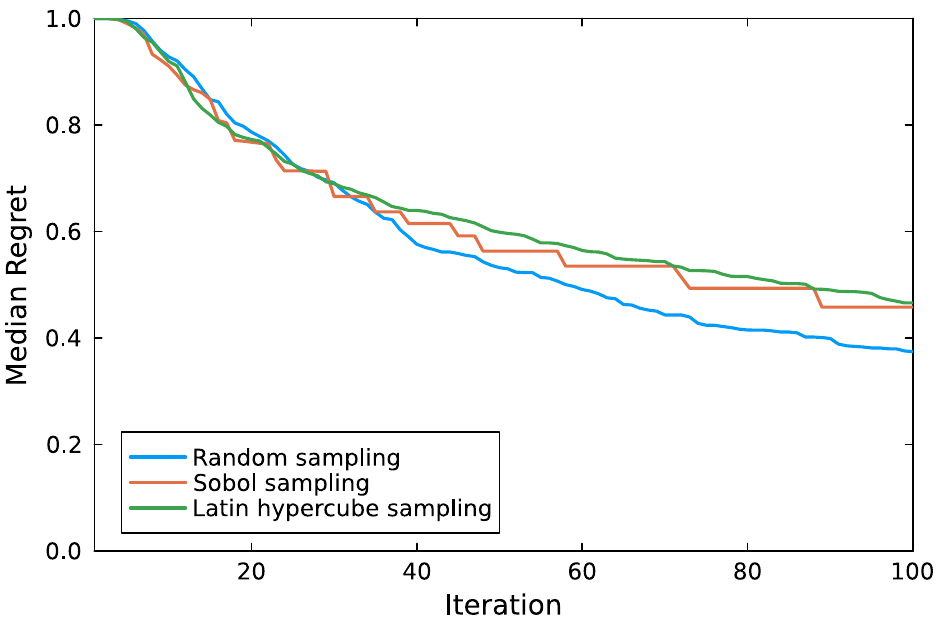}
  \caption{The regret of various pseudo-random sampling methods on the same phase as shown in Fig.~\ref{fig:benchmark}(a-b). The pseudo-random methods, including Sobol sequencing and Latin hypercube sampling, behave similarly to random sampling. All curves are from averaging over 500 synthetic tests.}
  \label{fig:pseudorandom}
\end{figure*}

Fig.~\ref{fig:pseudorandom} shows the comparison of the random sampling strategy with respect to two pseudo-random sampling methods. The reference benchmark dataset is identical to the one discussed in the main manuscript in Sec.~\ref{sec:synthetictests} and illustrated in Fig.~\ref{fig:benchmark}(a) and Fig.~\ref{fig:benchmark}(b).

\begin{figure*}[h!]
    \includegraphics[width=\linewidth]{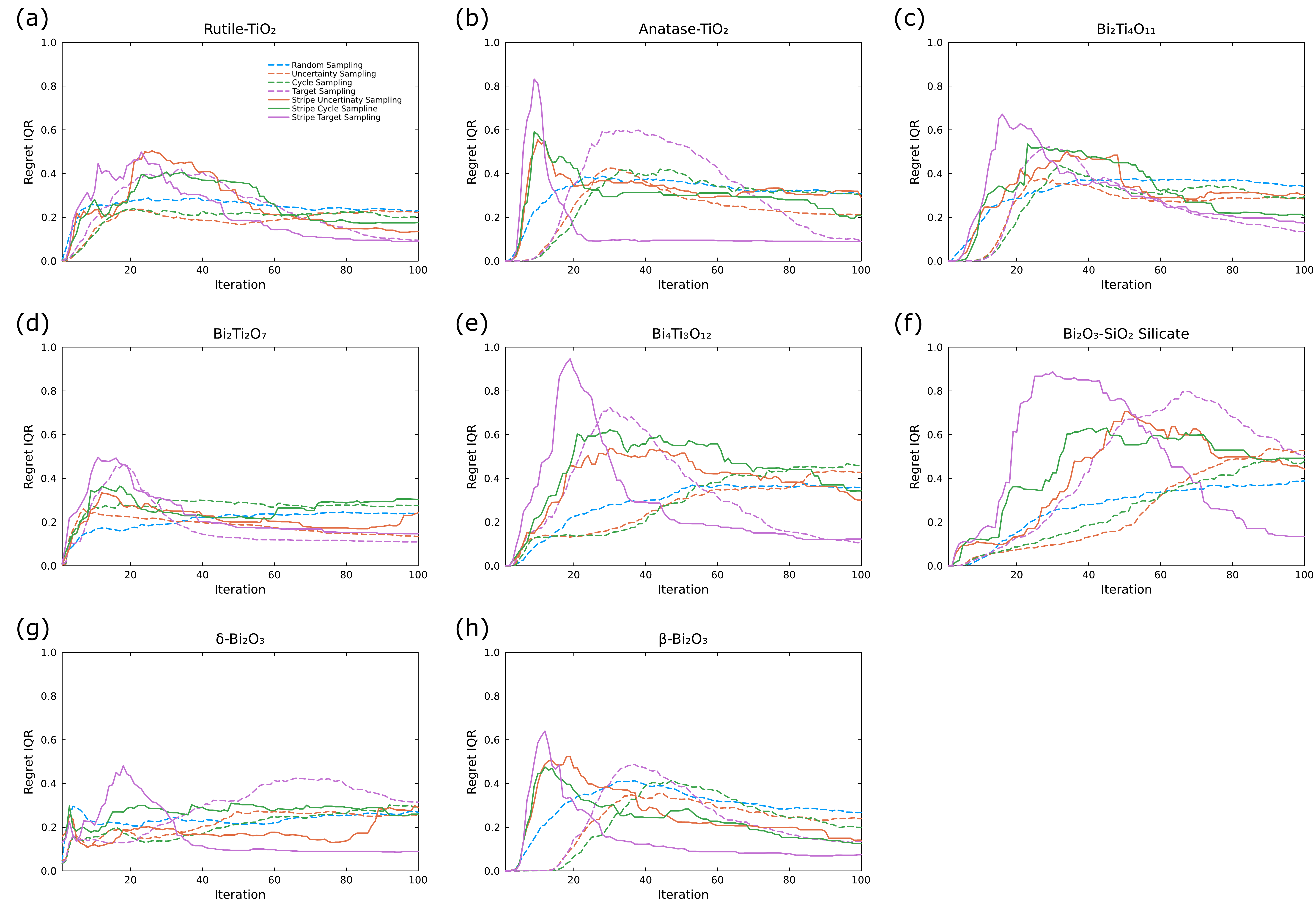}
  \caption{The IQR of regret curves for all phases involved in the benchmark dataset, including (a) Rutile-\ce{TiO2} (b) Anatase-\ce{TiO2} (c) \ce{Bi2Ti4O11} (d) \ce{Bi2Ti2O7} (e) \ce{Bi4Ti3O12} (f) \ce{Bi2O3}-\ce{SiO2} Silicate (g) \ce{\delta-Bi2O3} (h) \ce{\beta-Bi2O3}.
  In all of the phases, the trend of IQR is similar to anatase-\ce{TiO2}, which we show and discuss thoroughly in the main body of the manuscript, while the IQR peak height and iteration that peaked differ for different phases due to their unique distribution in the parameter space.}
  \label{fig:fulliqr}
\end{figure*}

Fig.~\ref{fig:fulliqr} shows the IQR of regret for all 8 benchmark datasets derived from the Bi-Ti-O system. Subfigure Fig.~\ref{fig:fulliqr}(b) is identical to the plot shown in Fig.~\ref{fig:benchmark}(b) of the main manuscript.

\FloatBarrier
\clearpage

\subsection{Detailed Discussions on the Enhancement and the Acceleration Factors\label{sec:enhancement-acceleration}}
We quantified the performance of each AL scheme by adapting the benchmarks proposed by Liang~\textit{et al.}~\cite{liang2021benchmarking}.
In particular, we considered the enhancement factor (EF) and the acceleration factor (AF).
The EF measures the performance gain for a fixed experimental budget, defined as the ratio of the maximum objective value achieved by the AL strategy to that achieved by random sampling after the same number of iterations.
An EF greater than one thus indicates that the AL strategy identifies superior solutions for a given cost.
In contrast, the acceleration factor (AF) quantifies the gain in experimental efficiency, calculated as the ratio of the number of iterations required by random sampling to the number required by the AL strategy to reach the same target objective value.
The AF therefore measures the speed-up, or the reduction in experimental cost, afforded by the intelligent sampling approach.
Detailed definitions of EF and AF are provided in Sec.~\ref{sec:metrics}.

The average EF curves over all eight phases in the parameter space of our AL strategies are shown and discussed in the main manuscript in Fig.~\ref{fig:benchmark}(c).
Similar to the results of the median regret, all AL strategies performed better than random sampling on average.
Cycle sampling outperformed uncertainty sampling in the long run, demonstrating the performance gain from exploiting phase information.
At the 100th iterations, the stripe cycle sampling and stripe target sampling were expected to find an objective value $1.8\times$ and $2\times$ as large as random sampling, demonstrating that AL based on phase information is useful for solving this optimization problem.

For comparison, the average AF over the eight phases for different AL strategies is shown in Fig.~\ref{fig:benchmark}(d).
From Fig.~\ref{fig:benchmark}(d), we can draw conclusions similar to those drawn from the EF. The higher AF of cycle sampling strategies compared to uncertainty sampling strategies, regardless of stripe or single-point variants, again demonstrates that the additional phase information enables better exploitation.
These AF curves also quantify the acceleration and resource savings in terms of experiments, which is crucial when resources are costly.
Hence, we can achieve the same results with three to six times fewer experiments, depending on how the scientist orchestrates the experiments, by applying AL and providing it with phase information to reason with.

\FloatBarrier
\clearpage

\subsection{Active Learning on the \ce{SnO_x} Suboxide System}
\label{sec:snot}
The Sn-O system encompasses several compounds and corresponding polymorphs that have significant applications across various sectors.
At ambient conditions, the most prevalent and stable form is \ce{SnO2} in its ground state rutile-type structure, which is recognized for its specialized utility as an electrode in electrochemical applications~\cite{allison_synthesis_2016}, as an n-type transparent semiconductor, and has been extensively explored for its gas sensing capabilities~\cite{das_sno2_2014}.

\begin{figure*}[b!]
\centering
  \includegraphics[width=\linewidth]{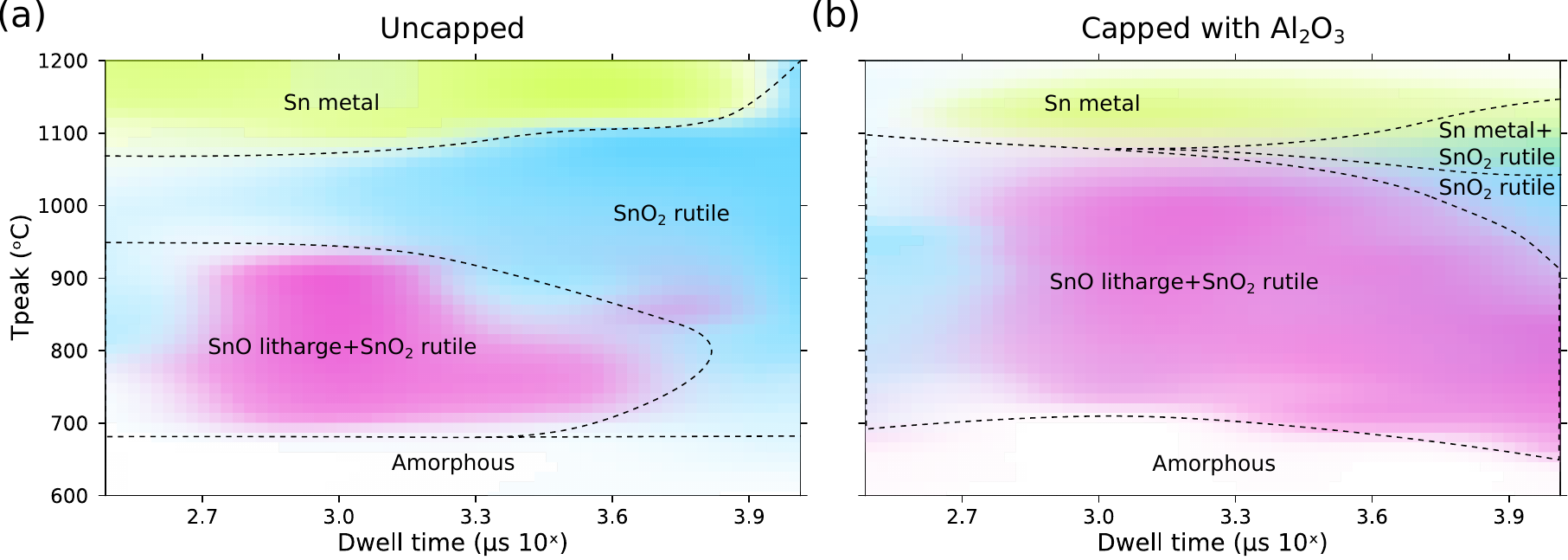}
  \caption{The experimentally determined processing phase diagrams of the \ce{SnO_x} system after SARA-H's converged AL cycles, both without and with a capping layer of \ce{Al2O3} in panels (a) and (b), respectively, aimed to investigate the role of atmospheric \ce{O2} during lg-LSA anneals. A total of 26 and 36 stripes are collected for the uncapped and capped portion of the phase diagram, respectively. Colors representing different phases are randomly selected from the HSV hue ring. The intensity of each region is proportional to the corresponding phase activation. The amorphous domain is represented as the uncolored region. The dashed black lines denote the presumed phase boundaries and are merely intended as a guide for the eye.}\label{fig:SNOT}
\end{figure*}

The second most prevalent phase in the Sn-O system is the litharge-type \ce{SnO}, which is especially valued for transparent semiconductor applications and demonstrates high p-type conductivity~\cite{caraveo-frescas_record_2013}.
Although it is not a thermodynamically stable phase, \ce{SnO} can be readily produced in bulk and thin film forms using conventional synthesis techniques such as RF-reactive sputtering; however, it decomposes into the rutile-type \ce{SnO2} and Sn metal upon heating to approximately 300~$^\circ$C.

Recently, Sutherland~\cite{sutherland_autonomous_2023} utilized lg-LSA in conjunction with synchrotron XRD analysis to systematically map the processing phase diagram of \ce{SnO_x} and study how phase formation can be affected by changing the oxidation state with respect to that of the initial film.
The authors investigated phase transformations in amorphous \ce{SnO2} and \ce{Sn2O3} suboxide thin film samples that were partially capped with an \ce{Al2O3} passivation layer to study the effect of atmospheric oxygen during lg-LSA processing.
The authors found drastically different phase behavior between suboxide and fully-oxidized films.
The suboxide film capped with \ce{Al2O3} displayed more retention of suboxide, demonstrating that significant transport of atmospheric oxygen into the film can occur in this system even in the ms time frame of an lg-LSA anneal.
For both the capped and uncapped sub-oxidized (and fully oxidized) samples, 105 lg-LSA anneals were performed under distinct processing conditions, thereby exhaustively sampling an equally spaced grid of the decadic logarithm of the dwell times $\tau$ between 550~$\mu$s and 10,000~$\mu$s, and peak temperatures $T_p$ between 500~$^\circ$C and 1400~$^\circ$C.
Independent of the oxidation state of the precursors, the authors reported the emergence of similar nucleating phases at relatively low temperatures (approximately 550~$^\circ$C).
These phases are characterized by broad diffraction peaks that deviate from any previously identified structures in the Sn-O system, undergoing a continuous second-order transformation and ultimately adopting the rutile-type \ce{SnO2} structure at elevated temperatures.
Less oxidized (uncapped and capped) precursor films predominantly established crystal structures of rutile-type \ce{SnO2} and SnO litharge.
While the SnO litharge structure persists for longer dwell times and higher temperatures in the capped film, the uncapped film shows more pronounced full oxidation. At high temperatures and extended dwell times, the \ce{SnO_x} phases undergo further decomposition and reduction reactions, resulting in the formation of metallic Sn.

Here, we reinvestigated the same sub-oxidized sample as a benchmark system to assess the performance of the SARA-H framework, both with and without the oxide capping layer. 
We populated the pool of candidate phases with those identified from exhaustive experiments, namely rutile \ce{SnO2}, litharge \ce{SnO}, and metallic Sn.
The primary goal here is to investigate how much oxidation can occur during lg-LSA and to demonstrate the advantages of using autonomous experiments with SARA-H over exhaustive sampling to achieve the same scientific insights into a system with fewer experimental iterations, thereby significantly reducing the time and other resources required to obtain scientific insight.

The converged phase diagrams from SARA-H's AL campaigns of the uncapped and capped samples are shown in Fig.~\ref{fig:SNOT}.
A comparison with the corresponding figures from the exhaustive sampling in Ref.~\onlinecite{sutherland_autonomous_2023} demonstrates excellent agreement.
While the phase transformations observed in both the uncapped and capped films are similar, some transformations exhibit different dwell and temperature dependencies.
The overall temperature-dependent transformations that we observe in both samples begin with the co-crystallization of \ce{SnO} and \ce{SnO2} at $T_p\approx 650^\circ$C, followed by complete oxidation to form \ce{SnO2}, and finally, decomposition into Sn metal at temperatures above $T_p\approx 1100^\circ$C.
However, in the uncapped sample, full oxidation from \ce{SnO} to \ce{SnO2} occurs at a temperature of $T_p\approx 950^\circ$C for short and intermediate dwell times and even right at the crystallization onset ($\approx 700 ^\circ$C) for long dwell times, due to atmospheric oxygen uptake during anneals, as shown in Fig.~\ref{fig:SNOT}(a).
On the other hand, the complete oxidation to \ce{SnO2} does not occur until reaching very high temperatures and long dwell times for the sample capped with \ce{Al2O3}, as seen in Fig.~\ref{fig:SNOT}(b).
Thus, the capping layer serves its purpose effectively, preventing atmospheric oxygen from migrating into the film and thereby inhibiting oxidizing the film to \ce{SnO2} across the majority of processing conditions.
Further, in-depth analysis of the effects of oxygen content and migration on the evolution of \ce{SnO_x} phases and their lattice strains will be published elsewhere.

\begin{figure*}[h!]
    \includegraphics[width=\linewidth]{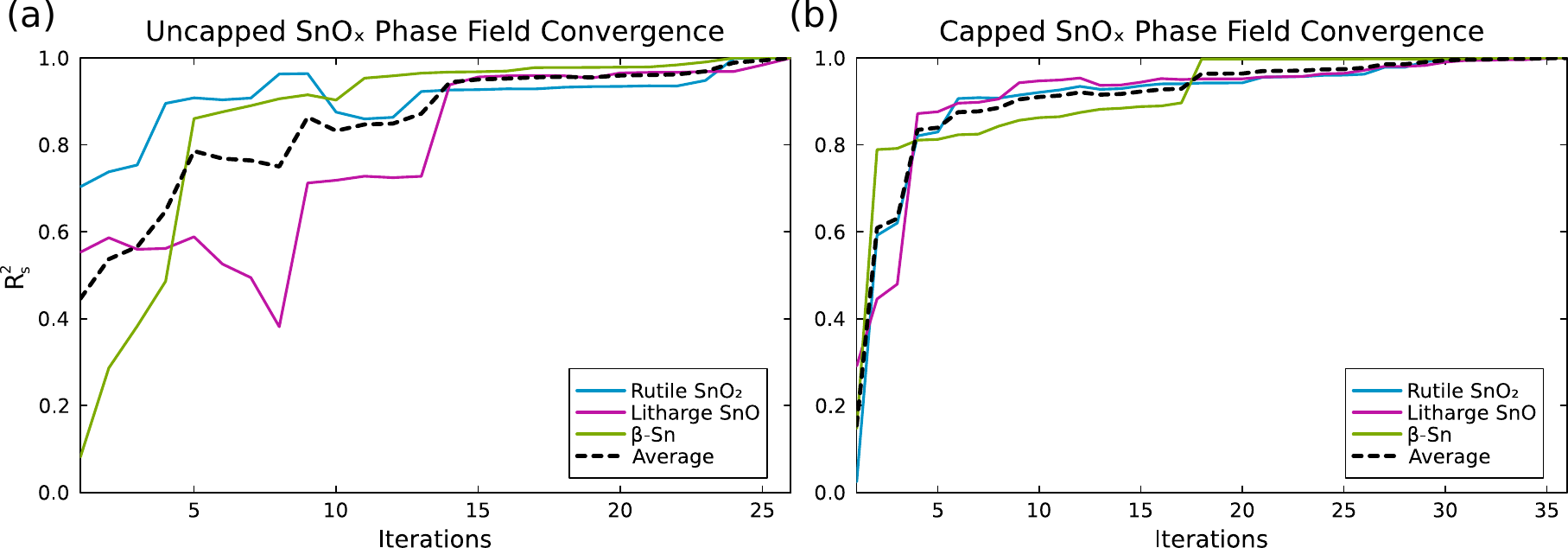}
  \caption{Panels (a) and (b) display the $R_s^2$ convergence as a function of the AL iterations for each of the observed phases in the uncapped and capped \ce{SnO_x} suboxide samples, respectively. The average values are shown as a dashed line.}
  \label{fig:SNOT_Convergence}
\end{figure*}

The learning curves of the uncapped and capped AL campaigns are shown in Fig.~\ref{fig:SNOT_Convergence}.
The total number of iterations is 26 for the uncapped film and 36 for the capped film.
For both systems, the average $R_s^2$ exceeds $R_s^2=0.8$ within fewer than 10 iterations, indicating that the \ce{SnO_x} phase diagram is notably easier for SARA-H to learn than \ce{Bi2O3}.
Compared to the 105 exhaustive experiments conducted by Sutherland~\cite{sutherland_autonomous_2023}, the savings in resources are greater than an order of magnitude.

\FloatBarrier
\clearpage

\subsection{Additional details on the Bi-Ti-O Material System}
\label{sec:bitio_review}
\subsubsection{Phases in the  Bi-Ti-O Material System}
The Bi-Ti-O system has a complex phase diagram which has been extensively studied in the past.
In addition to the phases discussed in the main manuscript, the following ternary phases have been investigated in the literature.

On the bismuth-rich side of the phase diagram, \ce{Bi12TiO20} was reported to form, most likely in the sillenite structure~\cite{speranskaya_system_1965,maier_notitle_1981}.
As the Ti fraction increases, Masuda~\textit{et al.}~\cite{masuda_crystal_1992} was the first to identify the \ce{Bi2Ti2O7} pyrochlore phase existing in the Bi-Ti-O system, which is a dielectric material with potential applications in relaxor ferroelectrics and exhibits high photocatalytic activity~\cite{zhou_synthesis_2006}.
It was later discovered that \ce{Bi2Ti2O7} decomposes into \ce{Bi2Ti4O11} and \ce{Bi4Ti3O12}~\cite{jiang_investigations_1999,toyoda_synthesis_1993,nakamura_preparation_1993} at elevated temperatures, the latter of which has also been touted as an effective photocatalyst.
Both \ce{Bi2Ti4O11} and \ce{Bi4Ti3O12} may coexist with \ce{TiO2}, potentially further improving the photocatalytic performance.

In addition to the intermediate compounds mentioned above, the effect of doping on \ce{TiO2} with additional elements has attracted interest in improving its properties, particularly by stabilizing the desired metastable anatase phase.
The impact of dopants on the transformation behavior of such metastable \ce{TiO2} phases has been extensively studied in the literature.
Hanaor~\textit{et al.}~\cite{hanaor_review_2011} provides a comprehensive review of the anatase-to-rutile transformation in \ce{TiO2} and summarizes the effect of cationic doping on the kinetic stability of anatase.
Based on empirical observations and the assumption of a substitutional solid solution, the substitution of Ti with small cations of valency lower than four is expected to accelerate the transition to rutile due to an increase in oxygen vacancies, which facilitates the rearrangement of the relatively large and rigid oxygen sublattice into rutile.
Conversely, large cations with high valency have been shown to inhibit the anatase-to-rutile transformation.
To the best of our knowledge, there are no reports on Bi doping of \ce{TiO2} and its effect on the stability of anatase. However, Hanaor~\textit{et al.}~\cite{hanaor_review_2011} offers a predictive tool to distinguish between inhibiting and promoting dopants, taking into account the Shannon ionic radii and atomic valency in Eq.~(6) of the report.
Based on this equation and the ionic \ce{Bi^{3+}} radius of 1.03~\AA\,~\cite{hanaor_review_2011}, we expect that doping \ce{TiO2} with Bi will inhibit the anatase-to-rutile transformation, a hypothesis we aim to test in this work.

Tab.~\ref{tab:ciflist} lists all phases considered in the Bi-Ti-O system.

\begin{table}[h!]
    \centering
    \begin{tabular}{|c|c|c|}
    \hline
        Composition & Structure/Phase & ICSD ID \\
        \hline
        \ce{Bi} &  A7 structure, $R\bar{3}m$ & 64703  \\
        \ce{SiBi2O5} & Silicate, $Cmc2_1$ & 30995  \\
        \ce{Bi2O3} & $\alpha$-\ce{Bi2O3}, $P2_1/c$ & 168567  \\
        \ce{Bi2O3} & $\beta$-\ce{Bi2O3}, $P\bar{4}2_1c$ & 189995  \\
        \ce{Bi2O3} & $\delta$-\ce{Bi2O3}, $Fm\bar{3}m$ & 189996  \\
        \ce{Bi_{7.68} Ti_{0.32} O_{12.16}} &  $P4_2/nmc$ & 83641  \\
        \ce{Bi4 Ti3 O12} &  Aurivillius, $Aba2$ & 16488  \\
        \ce{Bi2Ti2O7} & Pyrochlore, $Fd\bar{3}m$ & 99437  \\
        \ce{Bi2Ti4O11} & $\beta$-\ce{Bi2Ti4O11}, $C2/m$ & 79769  \\
        \ce{Bi36Ti32O} &  $P4/nmm$ & 256205  \\
        \ce{TiO2} & Anatase, $I4_1/amd$ & 154604  \\
        \ce{TiO2} & Rutile, $P4_2/mnm$ & 159915  \\
\hline
    \end{tabular}
    \caption{List of all 12 initial candidate phases for the AL campaign. The composition, structural details, and the ICSD ID number are provided in the three columns for each candidate.}
    \label{tab:ciflist}
\end{table}

\FloatBarrier
\clearpage
\subsubsection{Convergence Behavior for Bi-Ti-O}
\begin{figure*}[h!]
\includegraphics[width=0.7\linewidth]{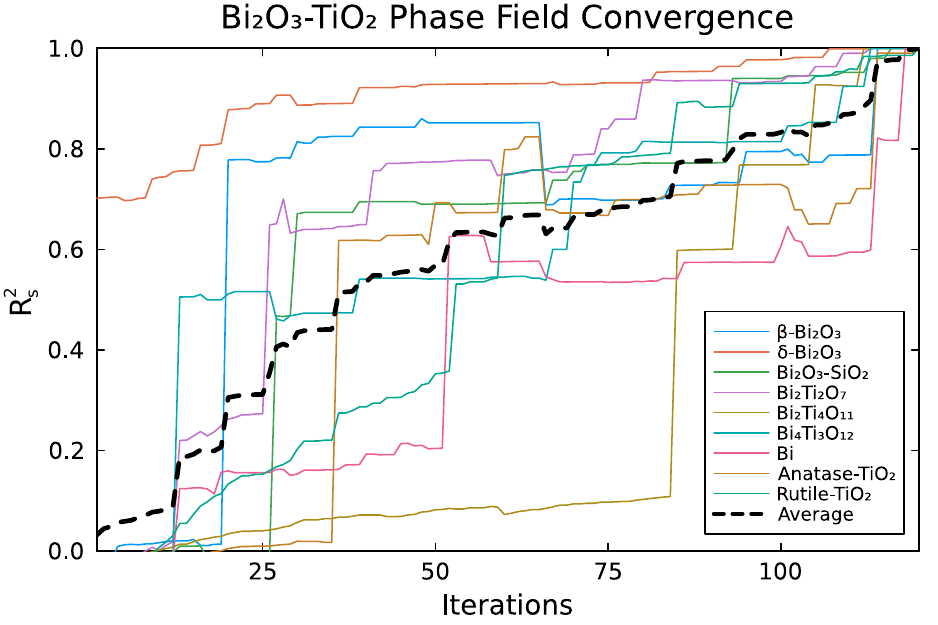}
  \caption{The $R_s^2$ convergence behaviors are presented as a function of the AL iterations for each of the observed phases in the Bi-Ti-O system. The average values are depicted as a dashed line.}
  \label{fig:BiTiO_Convergence}
\end{figure*}

This section contains the phase-resolved learning curves of SARA-H’s AL campaign for the Bi-Ti-O system in Fig.~\ref{fig:BiTiO_Convergence}.

\FloatBarrier

\clearpage

\clearpage
\subsubsection{Composition Spread Bi-Ti-O\label{sec:xrf}}
\begin{figure*}[h!]
\includegraphics[width=0.7\linewidth]{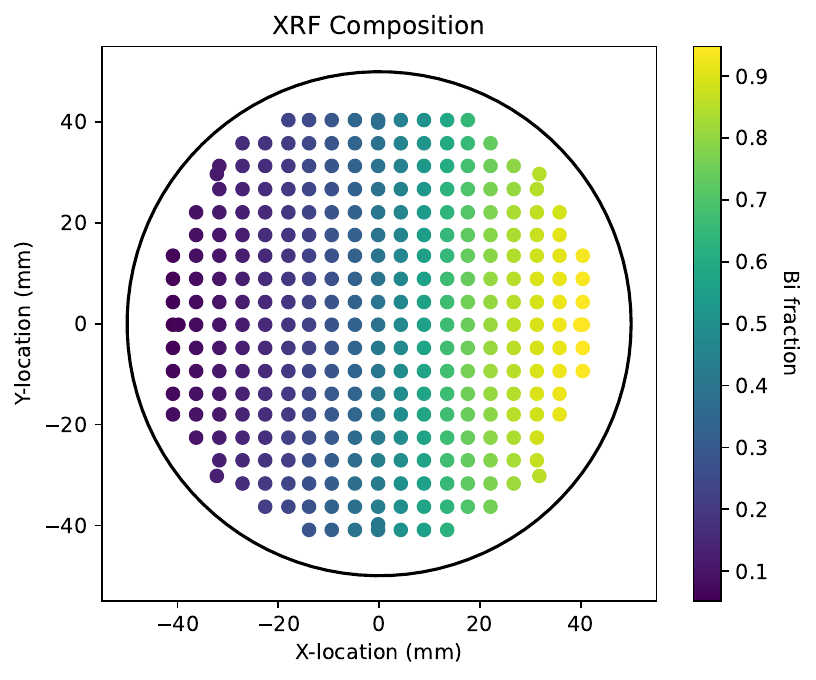}
  \caption{The Bi-fraction of the Bi-Ti-O composition spread as a function of the location on the Si substrate based on XRF measurements. The black circle signifies the boundary of the 100 mm Si wafer. For the AL campaigns, the compositional information was interpolated at each location at which an lg-LSA anneal was performed by fitting a smooth gaussian process regression model to the experimentally determined compositions shown in this figure. }
  \label{fig:BiTiO_XRF}
\end{figure*}

This section shows the locations on the 100~mm silicon wafer where the composition of binary spreads was evaluated using X-ray fluorescence (XRF) in Fig.~\ref{fig:BiTiO_XRF}.
\FloatBarrier
\clearpage

\subsubsection{Evolution of the Anatase-\ce{TiO2} Phase Field in the Bi-Ti-O System}
\begin{figure}[h!]
  \includegraphics[width=\linewidth]{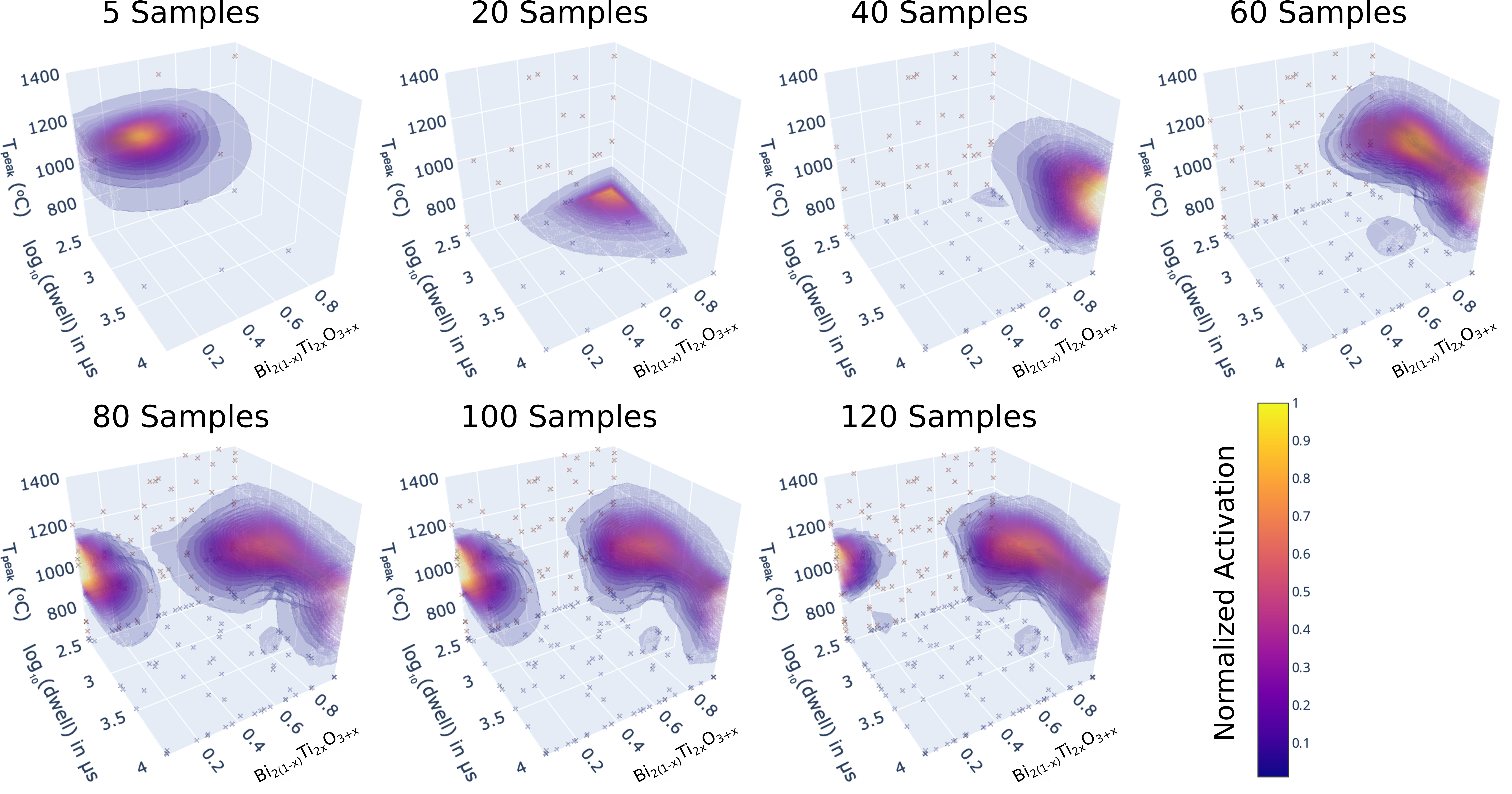}
  \caption{The normalized expected phase activation volumetric heat maps of anatase-\ce{TiO_2}, which each panel corresponding to the activation after the first $N=5, 20, 40, 60, 80, 100, 120$ iterations of SARA-H's AL cycles.
The orange crosses indicate the conditions $\{\textbf{c},T_p,\tau \}$ under which the actual anneals are conducted.
The activation is normalized so that the maximum activation of each graph is 1. 
For $N=5$ and $N=20$, very low probability densities are observed at conditions that do not actually contain the anatase-\ce{TiO_2}, as the phase field disappears in the later diagrams due to normalization.
At $N=40$ samples, SARA-H detects strong phase activation at high Ti content and quickly exploits this information to establish a large continuous distribution of anatase-\ce{TiO_2} along that Ti content.
The phase field converges after $N=80$ samples; however, a misclassification with a high probability creates an unrealistic phase field in the high Bi content region, which can be easily disregarded through human intervention in post-processing.
  }
  \label{fig:bo_process}
\end{figure}

This section contains the rendered phase volume of the anatase-\ce{TiO2} phase as a function of the number of anneals peformed during the AL campaign in Fig.~\ref{fig:bo_process}. 

\FloatBarrier

\clearpage

\subsubsection{Phase Field of Bi-Ti-O System Constructed from Autonomous Experiment Trial}
\begin{figure*}[h!]
\includegraphics[width=\linewidth]{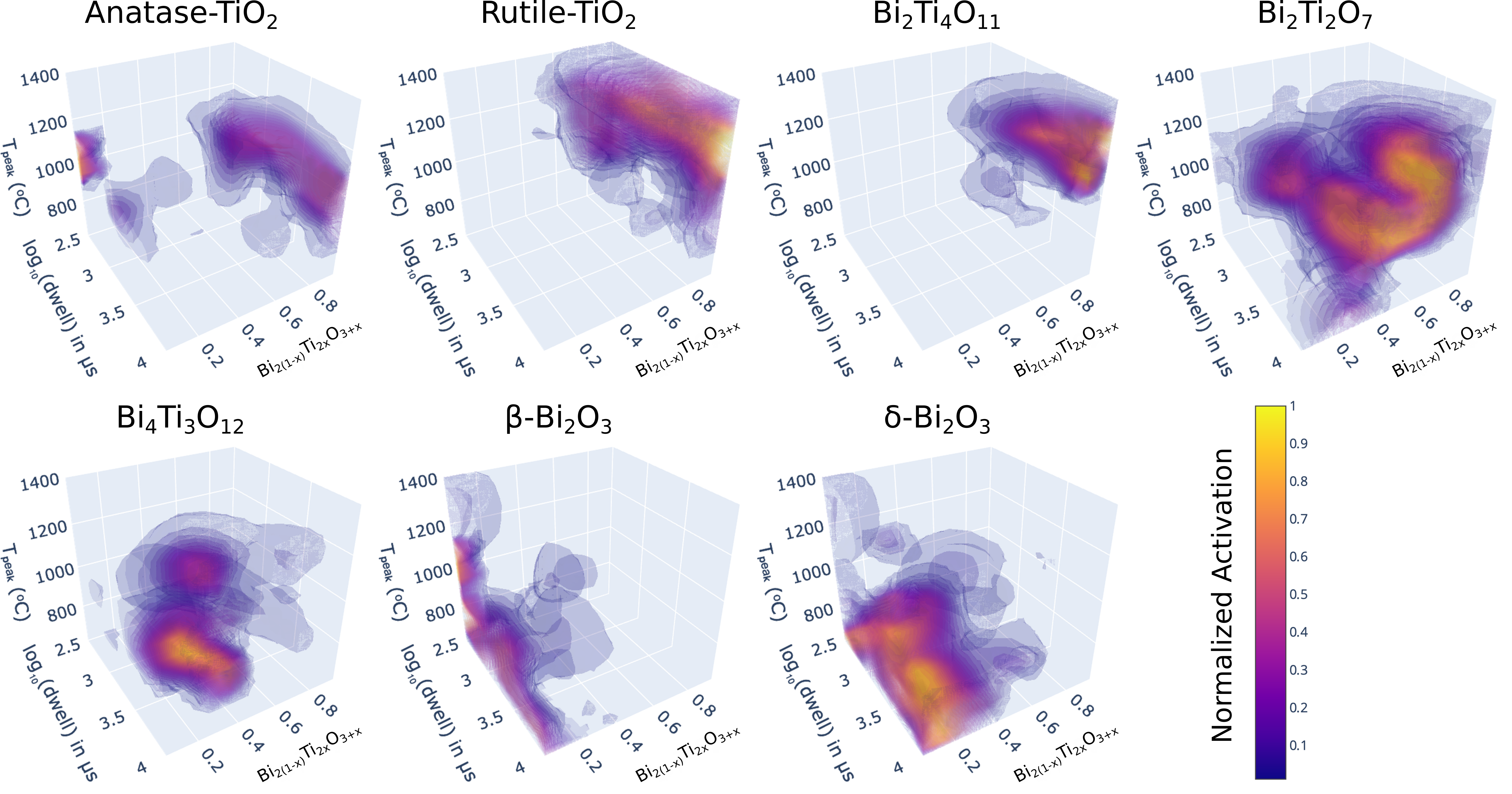}
  \caption{The autonomously constructed, normalized expected phase activation is presented as volumetric heat maps for each observed phase in the Bi-Ti-O system.
  The individual phase diagrams are generated after SARA-H's AL cycles have fully converged within 120 iterations.
  Similarly to Fig. \ref{fig:bo_process}, a misclassified domain at short dwell times and low Ti concentration in anatase-\ce{TiO2} results in a spurious blob in the phase field, which can be easily dismissed through human intervention or during post-processing.}
  \label{fig:AL_phase_diagrams}
\end{figure*}

This section contains the rendered phase volumes of all observed phases in the Bi-Ti-O system at the final stage of the AL campaign in Fig.~\ref{fig:AL_phase_diagrams}. 

\FloatBarrier

\clearpage

\subsection{Schematic Illustration of the lg-LSA and XRD Setup}
\begin{figure*}[h!]
\includegraphics[width=0.7\linewidth]{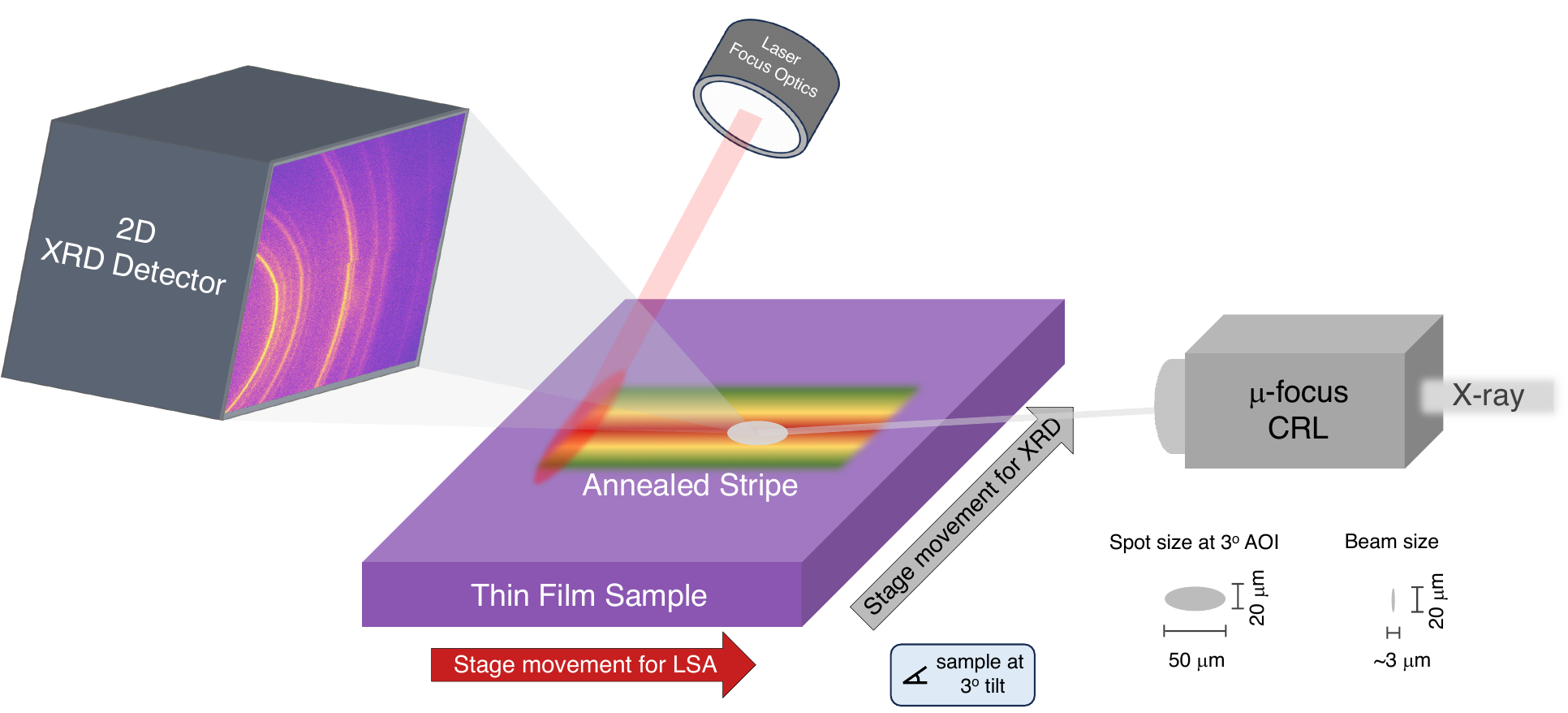}
  \caption{Schematic of the lg-LSA and XRD setup.
  lg-LSA was conducted by shining a focused laser on the thin film sample while the stage was moving and creates a processed stripe with different lateral position experiencing different temperature profile.
  Subsequent XRD mapping was conducted with the sample at a $3^\circ$ tilt.
  The X-ray produced by the synchrotron was focused by the roughly a $20 \times 3 \mu$m beam size and incident the film to create a $20 \times 60 \mu$m X-ray spot on the sample surface.
  A 2D detector was fixed in space and continuously collecting diffracted X-ray signals as the X-ray scans across the stripe.
  The relative position and the geometry of the setup can be calibrated by collecting diffraction patterns of a well-known material (e.g. gold) and retrieves such information using calibration packages like pyFAI.
  }
  \label{fig:setup}
\end{figure*}

This section contains the schematic illustration of the lg-LSA and XRD setup in Fig.~\ref{fig:setup}.
\FloatBarrier
\clearpage

\subsection{On the Timescale of the Autonomous Experimentation  \label{sec:timescales}}

In this section we break down the timescales required to conduct the autonomous experiments reported in the manuscript. Once a thin film sample is deposited, performing the autonomous anneals and stripe analysis to drive SARA-H's active learning cycles necessitates the following steps per iteration, along with their associated runtimes.

\begin{description}
    \item[lg-LSA] Each anneal of a single lg-LSA stripe requires approximately 15~s to complete. This time includes moving the stage to the designated starting position, accelerating the stage, performing the anneal, and decelerating the stage.
    \item[Microscopy Imaging] A microscopy image of every anneal strip is captured, which takes no more than 2 s.  
    \item[XRD] Each anneal is further analyzed by collecting 151 XRD data frames at 10~$\mu$m spacings across the stripe, with an integration time of 50~ms for each. In total, the data acquisition takes approximately 10~s.
    \item[2D-Integration] Each XRD frame is integrated into a 1D diffraction pattern and assembled into an XRD map, which is performed in parallel on a HPC cluster and takes around 15~s in total. 
    \item[CrystalShift and Bayesian Optimization] The analysis of the XRD map with CrystalShift and the evaluation of the phase map to propose the next anneal conditions takes between 20~s and 2~mins, depending on the number of candidate phases, the depth of the tree search, and the amount of aggregated phase data. The calculations are performed on an HPC cluster to parallelize the workload.
\end{description}

On a single 100~mm silicon wafer substrate, approximately 600 lg-LSA anneals can be performed. Based on above estimates, fully populating a thin film sample with 600 anneals can therefore be completed in approximately 20 hours.